\documentclass{aa501}
\usepackage{graphics}
\usepackage{epsfig}
\usepackage{amssymb}
\usepackage[english]{babel}
\newcommand{\be}{\begin{equation}}
\newcommand{\ee}{\end{equation}}
\newcommand{\bi}{\begin{itemize}}
\newcommand{\ei}{\end{itemize}}
\newcommand{\de}{\mbox{d}}

\newcommand{\mdot}{$\dot{M}~$} 
\newcommand{\mdisk}{$M_{disk}~$}

\title{The spectral energy distribution of self-gravitating protostellar disks}
\titlerunning{The SED of protostellar disks}
\author{G. Lodato \inst{1} \and G. Bertin \inst{1,2}}
\offprints{G.Lodato \email lodato@cibs.sns.it}
\institute{Scuola Normale Superiore, Piazza dei Cavalieri 7, I-56126 Pisa, 
Italy \and Universit\`a degli Studi di Milano, Dipartimento di Fisica, Via Celoria
16, I-20133 Milano, Italy}
\date{Received / Accepted}
\begin{document}

\abstract{
The long wavelength emission of protostellar objects is commonly attributed to a 
disk of gas and dust around the central protostar. In the first stages of disk 
accretion or in the case of high mass protostars, the disk mass is likely to be
sufficiently large, so that the disk self-gravity may have an impact on the dynamics
and the emission properties of the disk. In this paper we describe the spectral 
energy distribution (SED) produced by a simple, non-flaring, self-gravitating 
accretion disk model. Self-gravity is included in the calculation of the rotation 
curve of the disk and in the energy balance equation, as a term of effective heating
related to Jeans instability. In order to quantify in detail the requirements on the
mass of the disk and on the accretion rate posed on the models by realistic 
situations, we compare the SEDs produced by these models with the observed SEDs of a
small sample of well-studied protostellar objects. We find that relatively modest 
disks - even lighter than the central star - can lead to an interesting fit to the 
infrared SED of the FU Orionis objects considered, while in the case of T Tauri 
stars the required parameters fall outside the range suggested as acceptable by the 
general theoretical and observational scenario. On the basis of the present results,
we may conclude that the contribution of a self-gravitating disk is plausible in 
several cases (in particular, for FU Orionis objects) and that, in the standard 
irradiation dominated disk scenario, it would help softening the requirements 
encountered by Keplerian accretion models.
}

\maketitle

\section{Introduction}

It now seems well established that pre-main-sequence stars, such as T Tauri and FU 
Orionis objects, are generally surrounded by a disk of gas and dust. The traditional
evidence for such circumstellar disks comes from the excess luminosity found at long
wavelengths, far-infrared (Rydgren, Strom \& Strom \cite{strom}) and millimetric 
(Beckwith et al. \cite{beckwith}), indicating that this radiation comes from a 
region of relatively low temperature, far from the central star. 

The first attempt at trying to explain such excess infrared luminosity dates back to
the 70's (Lynden-Bell \& Pringle \cite{lynden}), when it was argued that it might 
be due to the release of gravitational binding energy from an accretion disk. In 
recent years, the general picture that has emerged is that, in most T Tauri systems,
the {\it mass accretion rate} is too low ($\dot{M}\sim 10^{-8}-10^{-7} M_{\odot}$
/yr) to account for the observed luminosity and that the observed long wavelength 
excess is likely to result from the reprocessing of the star-light from the outer 
disk (Hartmann \cite{hartmann}). Therefore, the accretion disk of T Tauri stars is 
generally thought to be ``passive''. In contrast, during FU Orionis outbursts, the 
accretion rate appears to be relatively high, so that here the disk can dissipate 
sufficient amounts of energy to produce the observed luminosity. In the construction
of dynamical models, a related issue is that of the total {\it disk mass}. Clues 
have indeed been gathered that the disk mass may be large in some protostellar 
systems (see Lay et al. \cite{lay}), especially in the earlier phases of the process
of star formation (Yorke \& Bodenheimer \cite{yorke}, for example, conclude from 
hydrodynamical simulations that the disk mass at the beginning of the accretion 
phase is comparable to that of the central object) and in the context of high-mass 
star formation, where observations point to the presence of massive disks (Cesaroni 
et al. \cite{cesaroni}). 

Standard accretion disk theory (Shakura \& Sunyaev \cite{shakura}) is based on the 
assumptions that the disk is optically thick and geometrically thin and that the 
power viscously dissipated in the disk $D=\nu_{\alpha}\sigma (r\Omega')^2/2$ is all 
radiated away. In the optically thick case, it can be shown that the spectral index 
$n=\de\mbox{Log}(\nu F_{\nu})/\de\mbox{Log}\nu$ in the infrared part of the spectrum
is related to the exponent $q$ in the surface temperature profile $T_s=T_{0}(r_0/r)
^q$, so that $n\simeq 4-2/q$ (Adams, Lada \& Shu \cite{shu}). For a Keplerian 
rotation curve one has $T_s(r)\propto r^{-3/4}$, leading to $L_{\nu}=\nu F_{\nu}
\propto\nu^{4/3}$, while for T Tauri stars one typically finds $L_{\nu}\propto
\nu^{2/3}$ (Kenyon \& Hartmann \cite{kenyon}). The difficulty becomes even more 
evident for the so-called ``flat-spectrum T Tauri'' (such as \object{T Tau} itself 
and \object{HL Tau}), which show a spectral index $n\approx 0$. In turn, if the 
rotation curve is flat, the surface temperature profile decreases more slowly 
$T_s(r)\propto r^{-1/2}$. In this latter case, the spectral index would be $n=0$, 
resembling that of flat-spectrum objects. Adams, Lada \& Shu (\cite{shu}) indeed 
proposed that the flat spectrum could arise from a disk with such a non-standard 
surface temperature profile, but argued that the observed disk mass in these systems
(although difficult to measure) is probably too low to produce substantial changes 
to the rotation curve and then suggested that the non-standard temperature profile 
should be ascribed to collective processes of angular momentum and energy transport 
associated with self-gravity induced instabilities.

In the absence of a convincing physical justification for non-standard descriptions 
of accretion disks, the attention has been turned to other ways of explaining the 
far-infrared spectra of T Tauri stars. As noted above, it is now commonly thought 
that in most of these systems the accretion rate is low and that the heating of the 
disk is primarily due to irradiation from the central star. The effect of 
irradiation is enhanced at large radii by the disk flaring, when the thickness of
the disk increases rapidly with radius. Models of the vertical structure of these 
flared disks have been constructed (Kenyon \& Hartmann \cite{kenyon2}; Chiang \& 
Goldreich \cite{chiang}; but concerns about stability have been expressed by 
Dullemond \cite{dullemond}). On the other hand, in the case of flat spectrum sources
and of most FU Orionis objects, flaring alone is ineffective and the far-infrared 
emission has been attributed to an infalling envelope of dust (Kenyon \& Hartmann 
\cite{kenyon3}; Calvet et al. \cite{calvet}).

The hypothesis that the flat long wavelength spectrum of protostellar objects arises
from a self-gravitating disk with flat rotation curve actually remains attractive 
from a physical point of view, especially for its simplicity. Unfortunately, only 
scattered efforts have been made in the direction of constructing self-gravitating 
models of accretion disks. The general ideas mentioned above have been revisited 
recently along with a new viscosity prescription for self-gravitating disks (Duschl,
Strittmatter \& Biermann \cite{duschl}), but little or no physical justification has
been provided for the desired radial surface density profile. In view of this 
discussion, it seems natural to consider and to put to a test the accretion disk 
models that we have constructed recently (Bertin \cite{bertin}; Bertin \& Lodato 
\cite{lodato}, hereafter BL99), which include the gravitational fields of both the 
central star and the disk. Interestingly, these models show that the rotation curve 
of a self-regulated accretion disk is Keplerian in the inner disk and flat in the 
outer disk, corresponding to a surface density $\sigma\propto r^{-1}$ (which, 
incidentally, has been observed in some cases: e.g., \object{TW Hya}, Wilner et al. 
\cite{wilner}, and \object{HH 30}, Burrows et al. \cite{burrows}), the transition 
taking place at a radius $r_s\approx(GM_{\star}/8)(G\dot{M}/2\alpha)^{-2/3}$ (using 
standard notation; see Section \ref{model}). Deviations from Keplerian rotation 
occur even for disk masses $M_{disk}\lesssim M_{\star}$, which could be a plausible 
range even for some T Tauri disks. Based on these models of self-regulated 
accretion, {\it it is now possible to quantify in detail what would be the parameter
requirements} needed to fit the infrared spectra of protostellar objects. In this 
respect, our results will turn out to be quite different (see discussion in Sect. 
\ref{simple}) from earlier estimates (e.g., see Kenyon \& Hartmann \cite{kenyon2}; 
Shu, Adams \& Lizano \cite{lizano}) that did not include the important ingredient of
self-regulation, which is likely to change significantly the energy balance 
equations in the self-gravitating part of the disk (see also discussion in Sect. 
\ref{energy}).

In this paper we describe the spectral energy distribution produced by a class of 
self-regulated, self-gravitating accretion disks and discuss its dependence on the 
various parameters involved and the differences from the non-self-gravitating case. 
In addition, we determine the physical parameters needed to fit the SEDs of 
realistic protostellar objects, by referring to a small sample of well-studied FU 
Orionis systems and T Tauri stars. A general theory of FU Orionis and T Tauri 
systems, or a comprehensive picture of the individual objects picked here to test 
the pure self-gravitating model against observed SEDs, is beyond the goals of this 
paper. The models considered are extremely simple and do not address the issue of 
the physical ingredients that determine the long-term evolution of the system (in 
particular, the physical ingredients responsible for the outburst phase of the FU 
Orionis systems): they are taken to match current ``snapshots'' of such evolving 
systems, assuming that the steady-state equations are temporarily adequate to 
describe the observed situation. 

The quality of the fit to the available data for the long wavelength spectral energy
distribution is very good. In the case of FU Orionis systems the required accretion 
rate is of the order of $\dot{M}\approx 10^{-5}-10^{-4}M_{\odot}/$yr and the implied
disk mass is of the order of one solar mass. The implied mass accretion rate for T 
Tauri stars is found to be $\dot{M}\approx 10^{-7}-10^{-6}M_{\odot}/$yr. The choice 
of the outer radius of the disk determines the mass of the disk relative to that of 
the central star. The estimates, although uncertain, point to rather massive disks 
($M_{disk}\approx 0.3M_{\odot}$). In the case of FU Orionis systems the mass 
accretion rates are in reasonable agreement with the currently accepted estimates, 
while the disk masses are higher than the commonly assumed disk masses in these 
systems. For T Tauri stars, instead, the required parameters tend to fall outside 
the range suggested as acceptable by the general theoretical and observational 
scenario. In any case, the contribution of self-gravity may soften the demands on 
models that attempt a description without the use of the disk self-gravity.

The paper is organized as follows. In Section \ref{model} we describe the model of 
self-gravitating, self-regulated accretion disk that we adopt here; in Section 
\ref{seds} we describe the spectral energy distributions derived on the basis of our
model and discuss their general properties; in Section \ref{fit} we consider a 
sample of FU Orionis objects and fit their SED with the self-gravitating disk model,
deriving the physical properties of the disks inferred on the basis of our model, 
and discuss the results obtained; in Section \ref{ttau} we point out the 
difficulties met when trying to adopt a similar approach for the case of T Tauri 
stars; in Section \ref{conclusion} we draw our conclusions.

\section{Self-regulated accretion disks}
\label{model}

\subsection{Disk model}

We refer to a self-gravitating accretion disk model that is only partially
self-regulated (BL99), so that the relevant $Q$-profile is of the form:
\begin{equation}
\label{qprofile}
Q=\frac{c_s\kappa}{\pi G\sigma}=\bar{Q}\left[1+\left(\frac{r}{r_Q}\right)^{-9/8}
\exp(-r/r_Q)\right],
\end{equation}
where $c_s$ is the {\it effective} thermal speed, $\kappa$ is the epicyclic 
frequency, and $\sigma$ is the surface density of the disk. In this way the 
stability parameter $Q$ is much higher than unity in the inner disk ($r\lesssim 
r_Q$) and becomes constant ($Q\sim\bar{Q}\approx 1$) in the outer disk. While in the
context of galactic dynamics the scale $r_Q$ is usually associated with the bulge 
size and, in the context of AGN, some efforts have been made in the description of 
the transition from a non-self-gravitating to a self-gravitating accretion disk 
(Bardou et al. \cite{bardou}), no clear-cut prescription for the onset of 
self-regulation in protostellar disks is available. We have thus decided to take 
$x_Q=r_Q/r_s$ (for the definition of $r_s$, see Eq. (\ref{rs}) below) as a free
parameter of our model. The scale $r_s$ is the scale that defines the transition 
from Keplerian to flat rotation curve:
\begin{equation}
\label{rs}
r_s=2GM_{\star}\left(\frac{\bar{Q}}{4}\right)^2\left(\frac{G\dot{M}}{2\alpha}
\right)^{-2/3}.
\end{equation}
Here $M_{\star}$ is the mass of the central object, $\dot{M}$ the accretion rate, 
and $\alpha$ the viscosity parameter of the Shakura \& Sunyaev (\cite{shakura}) 
prescription
\begin{equation}
\label{shak}
\nu_{\alpha}\sigma=\alpha c_sh\sigma=\frac{\alpha c_s^3}{\pi G}\tilde{h} (Q,
\Omega/\kappa);
\end{equation}
in the last equation we calculate the disk thickness as 
$h=(c_s^2/\pi G\sigma)\tilde{h}$, with:
\begin{equation}
\label{htilde}
\tilde{h}=\frac{\pi}{4Q^2\left[2\Omega^2/\kappa^2-1\right]}
\left[\sqrt{1+\frac{8}{\pi}Q^2\left(\frac{2\Omega^2}{\kappa^2}-1\right)}-1
\right].
\end{equation}
The last expression gives a useful interpolation formula for the thickness of 
the disk in the case when the contributions of the star and of the disk are
both taken into account in the vertical gravitational field (see BL99, Appendix).

In the following, for simplicity we assume that the conservation law for the 
angular momentum is written as:
\begin{equation}
\label{consang}
G\dot{M}r^2\Omega+2\alpha c_s^3\tilde{h}r^3\frac{\de\Omega}{\de r}\simeq 0
\end{equation}
and thus avoid the relatively complex behavior that is associated with a net 
transport of angular momentum ($\dot{J}\neq 0$; calculated by BL99). Here we may 
also recall that Popham et al. (\cite{popham}) find better agreement with the 
observational data when their boundary layer models for FU Orionis systems are taken
with a negligible value of $\dot{J}$.

Finally, the rotation curve is computed by solving the full Poisson equation,
including the contribution of both the central star and the disk:
\begin{equation}
\label{rotcurve0}
V^2=\frac{GM_{\star}}{r}+r\frac{\mbox{d}\Phi_{\sigma}}{\mbox{d}r},
\end{equation}
where the gravitational field of the disk $\de\Phi_{\sigma}/\de r$ is given in 
Eq. (4) of BL99. In practice, a useful way to express the rotation curve is by 
scaling $V^2$ to its value in the outer disk:
\begin{equation}
\label{rotcurve}
V^2=\frac{1}{2}\left(\frac{4}{\bar{Q}}\right)^2
\left(\frac{G\dot{M}}{2\alpha}\right)^{2/3}\phi^2,
\end{equation}
with $\phi^2$ given in Eq. (13) of BL99. 

Equation (\ref{consang}) gives an expression for the profile $c_s(r)$ in terms of
$\Omega(r)$ and $Q(r)$; the latter functions are available from Eq. (\ref{rotcurve})
and Eq. (\ref{qprofile}). Similarly, Eq. (\ref{shak}) provides the profile of the 
quantity $\nu_{\alpha}\sigma$. Therefore, we have an explicit expression, dependent 
on few parameters (such as $\alpha$, $\dot{M}$, $\bar{Q}$), for the viscous 
dissipation rate $D(r)=\nu_{\alpha}\sigma(r\Omega')^2/2$. In the standard theory 
this would be sufficient to specify the surface temperature profile $T_s(r)$ 
associated with the optically thick emission of the disk. Here, as described in a 
separate article (Bertin \& Lodato \cite{lodato2}, hereafter BL01; see also 
additional remarks in Sect. \ref{energy} below), it is 
appropriate to consider an additional heating term in the energy balance equation, 
so that the surface temperature profile is determined by:
\begin{equation}
\label{surftemp}
\sigma_BT_s^4=\frac{1}{2}\nu_{\alpha}\sigma(r\Omega')^2+\frac{g(Q)}{Q}\frac
{c_s^3}{2\pi G}\kappa\Omega,
\end{equation}
with $g(Q)=(\bar{Q}/Q)^m$, $m$ a large number (we will take $m=20$), and where 
$\sigma_B$ is the Stefan-Boltzmann constant. The additional Jeans-related term 
ensures that the process of self-regulation properly occurs. Note that in Eq. 
(\ref{surftemp}) we neglect the contribution to the heating of the disk associated 
with the reprocessing of the light emitted by the central star, which is commonly 
thought to be the key heating source for T Tauri disks, especially if the outer 
disk is ``flared''; we will discuss the effects of disk irradiation below (Sect.
\ref{irradiation}). In any case, our model of self-gravitating disk has negligible 
flaring (see BL99).

As a final remark, we note that the assumption that the emission of the disk is 
optically thick is not conclusive, primarily because of the uncertainties in dust
opacity. In this respect, our recent analysis of the energy budget in self-regulated
disks (BL01), although it was focused on the different context of AGN, has pointed 
out that self-gravitating disks may be optically thin. Here, for simplicity and for 
a direct comparison with other studies, we consider models characterized by 
optically thick emission.

The notation in the present paper is slightly different with respect to earlier
articles (in relation to the use of $c_s$ and $\nu_{\alpha}$), in order to avoid
confusion with the universal notation for some physical quantities (speed of light 
and frequency).

\subsection{Self-gravity, viscosity, and the energy budget}
\label{energy}

Where does the energy of the Jeans-related term in Eq. (\ref{surftemp}) come from? 
Why is the heating associated with instabilities not necessarily related to 
$\dot{M}$ (through $\nu_{\alpha} \sigma$)? The reader is referred to a separate 
article (BL01) devoted to introducing and justifying our approach. Here we take this
opportunity to add some additional comments that might help answer questions such as
those posed above.
 
Earlier attempts at incorporating the role of collective effects associated with the
disk self-gravity generally focused on the problem of angular momentum transport and
thus on a modification of the viscosity prescription; these approaches (in 
particular, see Lin \& Pringle \cite{lin}, \cite{lin2}; Bardou et al. 
\cite{bardou}) often realized the need for a mechanism of self-regulation, locally
sensitive to the value of $Q$. In turn, numerical simulations (e.g., see Laughlin \&
Bodenheimer \cite{laughlin}, Laughlin \& R\'{o}\.{z}yczka \cite{laughlin2}) have
demonstrated that spiral density waves can indeed transport angular momentum so as 
to act as an ``effective viscosity''. In this respect, we agree that gravitational 
instabilities may have the welcome role, in the outer parts of accretion disks, of 
contributing significantly to the effective viscosity of the disk, and hence 
directly to the accretion rate $\dot{M}$.
 
In the approach that we have followed (starting with Bertin \cite{bertin}) we have
argued that the process of self-regulation, which is expected to maintain the outer 
disk at values of $Q$ of order unity, should result from a modification of the 
energy equations, so as to take into account the heating induced by gravitational 
instabilities. The processes involved are complex (see Sect. 3 in BL99 and Sect. 3 
in BL01). We should not expect that all the heating rate should be proportional to
the angular momentum transport (as the ideal fluid equations with a viscous term 
generally imply), especially since axisymmetric instabilities are known also to take
place and gravitational heating is observed even in collisionless disks (e.g., see 
the simulations by Hohl \cite{hohl} and Hohl \cite{hohl2}). Therefore, we have 
thought it appropriate to separate the two issues, of the angular momentum and of 
the energy transport.
 
In some extremely simple models (e.g., those considered by N-body simulations of 
collisionless disks that bring out the effects of evolution induced by Jeans-related
instabilities) one may argue, for example on the basis of the virial constraint, 
that if, globally speaking, energy is lost by the system, it must ultimately come
from the gravitational potential well. Locally, the statement need not be true, 
given the facts that some energy may be redistributed across the system through 
global instabilities (see remarks by Adams et al. \cite{adams2}) and that 
gravitational energy has little local meaning. Thus, if we refer to Eq. 
(\ref{surftemp}), this local relation does neither contradict nor confirm the 
requirements of the global energy budget. (Note that, much in line with other 
descriptions of accretion disks, we do not make explicit use of a horizontal energy 
transport equation.)

This general message, that the entire system cooperates globally to produce effects 
that show up differently in different parts of the disk, is often apparent from the 
results of a variety of numerical simulations (e.g., see Laughlin \& 
R\'{o}\.{z}yczka \cite{laughlin2}; Nelson et al. \cite{nelson2}). In particular, 
several numerical studies have addressed directly the issue of the energy budget 
when dynamical instabilities are involved (for some recent studies, see Pickett et 
al. \cite{pickett1}, \cite{pickett2}, and Nelson et al. \cite{nelson1}, 
\cite{nelson2}).
 
In addition, the disks that mediate the accretion process are actually not isolated:
in fact, for the young stellar objects considered in this paper there is a central 
star with which the disk connects through a possibly hot and radiatively inefficient
inner boundary layer (for FU Orionis objects, see Popham et al. \cite{popham}) and 
the disk is embedded in some environment from which the protostellar cloud has 
originated. These components, external to our set of model equations, represent a
significant energy reservoir.

In conclusion, we are not yet ready to produce an explicit set of closed equations 
to describe the overall energy budget. In fact, to do so we should not only describe
the global processes involved in the energy transport and the physical mechanisms 
active at the boundaries, but also face in detail the way we visualize an extremely 
complex system, in which the relevant heating and cooling terms may not be described
by a simple one-fluid model (for example, for what concerns possible inelastic
collisions between clumps). In this respect, it is interesting to see that in many
numerical experiments relatively complex systems are generated, with transient 
interacting structures (Pickett et al. \cite{pickett2}). On the other hand, the 
simpler heuristic approach that we have undertaken (in line with the spirit of the 
Shakura \& Sunyaev prescription) can overcome this difficulty and provide insight 
into the properties of self-gravitating disks.

While more work is certainly needed to set up a satisfactory global model, for the 
time being we proceed with this set of model equations that appear to possess most 
of the necessary ingredients required to describe a truly self-gravitating disk.

\section{General properties of the spectral energy distribution of active,
self-gravitating accretion disks}
\label{seds}
\subsection{Spectral energy distribution for disk dominated cases} 
\label{spectral}
From Eqs. (\ref{shak}), (\ref{consang}), and (\ref{surftemp}), it is useful to 
define a temperature scale $T_0$ (with frequency $\nu_0=kT_0/2\pi\hbar$) given by:
\begin{equation}
\label{eq:deft0}
T_0^4=\frac{\dot{M}}{2\pi\sigma_B\alpha^2}\left(\frac{2}{\bar{Q}}\right)^6
\left(\frac{\dot{M}}{M_{\star}}\right)^2,
\end{equation}
with a dimensionless temperature $\hat{T}=T_s/T_0$ and a dimensionless frequency 
$\hat{\nu}=2\pi\hbar\nu/kT_0=\nu/\nu_0$. 

For a disk inclined at an angle $\theta$ with respect to the observer's line of 
sight, at a distance $D_0$, the luminosity is provided by integration of the Planck
spectrum at the local surface temperature $T_s(r)$ between the inner radius $r_{in}$
and the outer boundary radius $r_{out}$ of the disk:
\begin{equation}
\label{diskflux}
\nu F_{\nu}=L_{\nu}=\frac{\cos\theta}{D_0^2}\int_{r_{in}}^{r_{out}}\frac{4\pi
\hbar\nu^4}{c^2}\frac{2\pi r\de r}{e^{2\pi\hbar\nu/kT_s(r)}-1}~.
\end{equation}

Using the dimensionless quantities defined above, the resulting spectrum can be 
written as:
\begin{equation}
4\pi D_0^2\nu F_{\nu}=L_D\hat{\nu}^4\int_{x_{in}}^{x_{out}}\frac{x\de x}
{e^{\hat{\nu}/\hat{T}(x)}-1},
\end{equation}
where:
\begin{equation}
\label{eq:defld}
L_D=\frac{32\pi^3r_s^2\hbar\nu_0^4}{c^2}\cos\theta=\frac{240\cos\theta}{\pi^4
\bar{Q}^2}\dot{M}\left(\frac{G\dot{M}}{2\alpha}\right)^{2/3}~.
\end{equation}
Note that the scale $L_D$ of the disk luminosity does not depend on $M_{\star}$. 

The spectrum of the disk is then completely defined by two scale parameters $L_D$ 
and $T_0$ and by four dimensionless quantities: the dimensionless inner and outer 
radii of the disk, $x_{in}=r_{in}/r_s$ and $x_{out}=r_{out}/r_s$, the viscosity 
parameter $\alpha$, and the dimensionless radius $x_Q$, defining the inner 
boundary of the self-regulated region (see Eq. (\ref{qprofile})). The two geometry
independent scale parameters, $L_D/\cos\theta$ and $T_0$ can be traced back to the 
values of $\dot{M}$ and $M_{\star}$. 

\subsection{Effects generated by variation of the key parameters}
\label{parameters}

In this subsection we briefly describe the changes in the SED induced by changing 
some key parameters. We start from a specific reference model, 
for which the relevant parameters are listed in Table \ref{tab:reference}. For such
reference model the scalelength $r_s$ turns out to be $r_s\approx 14$AU, so that the
scalelength beyond which the disk is self-regulated is $r_Q=x_Qr_s\approx 5.6$AU.
 
\begin{figure}[hbt!]

\centerline{\epsfig{figure=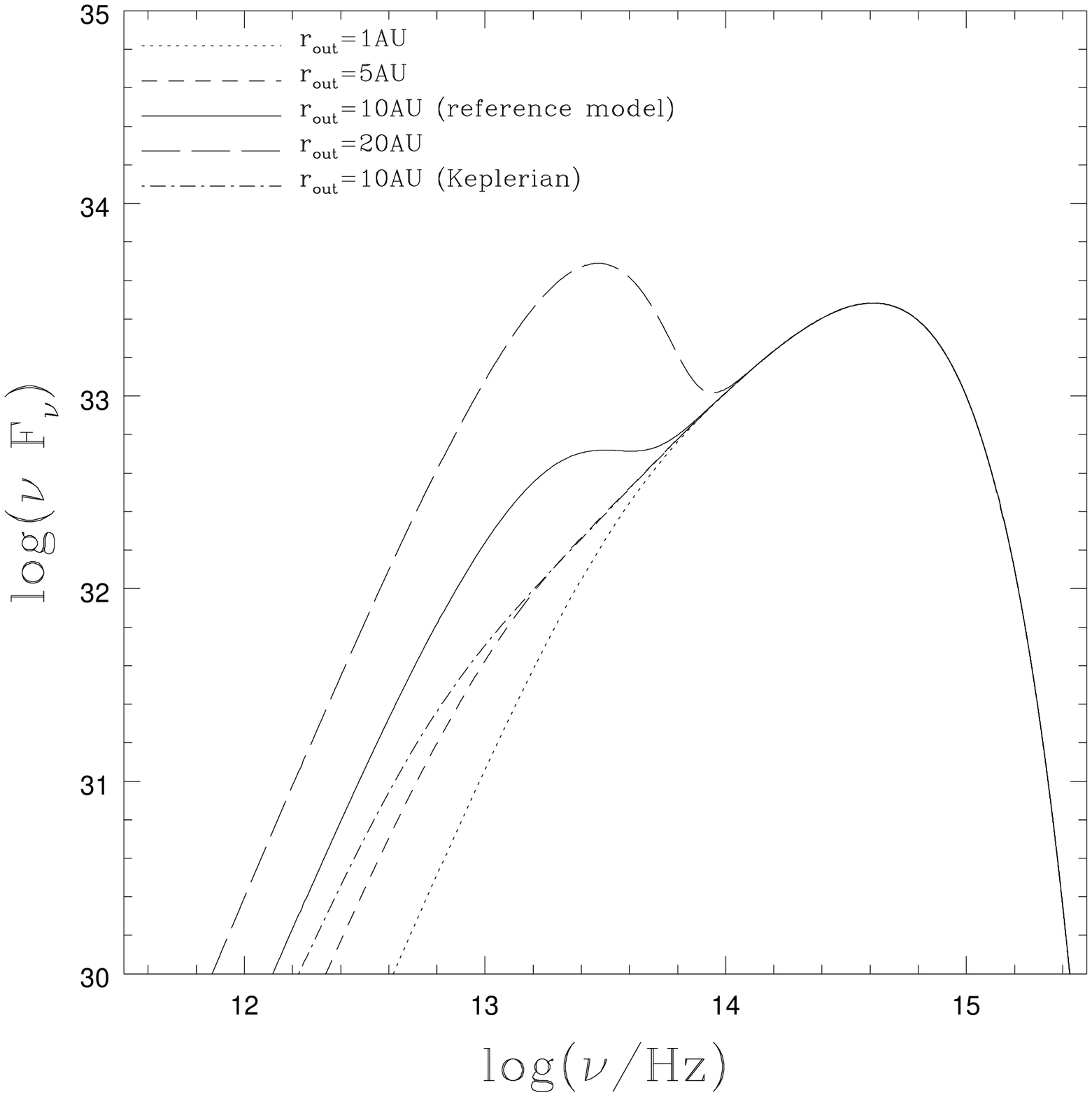,height=8cm,width=8cm}}
\caption{\small{Shape of the spectral energy distribution for different values of
the outer radius $r_{out}$. The quantity $\nu F_{\nu}$ is in arbitrary units.}}
  \label{fig:rout}
\end{figure}

Figure \ref{fig:rout} illustrates the effects of varying the outer radius of the 
disk. As expected, increasing the outer radius leads to a higher infrared 
excess. For comparison, Fig. \ref{fig:rout} also shows a model indicated by 
``Keplerian''. This model is computed with the same physical parameters as for the
reference self-gravitating model (i.e., the same $\dot{M}$, the same inner and
outer radii, etc.), but {\it as if} self-gravity were turned off, i.e. by taking a
strictly Keplerian rotation curve and by neglecting the additional term proportional
to $g(Q)$ in Eq. (\ref{surftemp}). The self-gravitating model clearly shows a much 
flatter SED in the far infrared. In contrast, for $r_{out}=1$AU the two models 
(self-gravitating and Keplerian) would produce practically coincident SEDs. This is 
because for $r_{out}=1$AU the outer radius of the disk is smaller than $r_Q$, so 
that the disk has not become self-gravitating in this case.

\begin{table}[hbt!]
\begin{center}
\begin{tabular}{ccc}
\hline
\hline
 Geometry                 &                               &  $\cos\theta=1$       \\
 Accretion parameters     & $\dot{M}=10^{-5}M_{\sun}/$yr  &  $M_{\star}=1M_{\sun}$\\
 Size of the disk         & $r_{in}=5R_{\sun}$            &   $r_{out}=10$AU     \\
 Viscosity and self-regulation &  $\alpha=10^{-3}$        &   $x_Q=0.4$           \\

\hline

\end{tabular}
\end{center}
\caption{\small{Parameters of the reference model}}
\label{tab:reference}
\end{table}

\begin{figure}[hbt!]

\centerline{\epsfig{figure=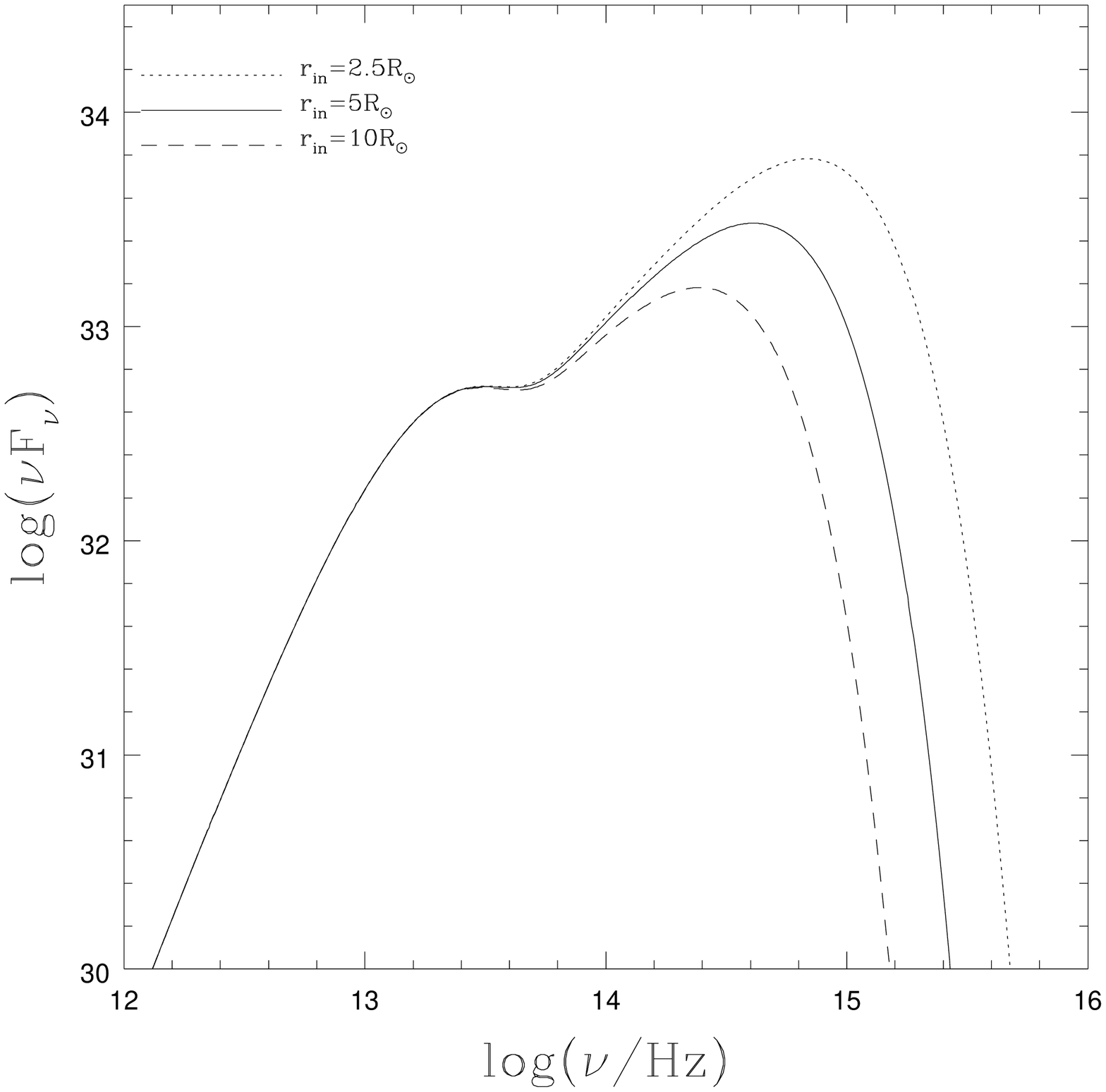,height=8cm,width=8cm}}
\caption{\small{Shape of the spectral energy distribution for different values of
the inner radius $r_{in}$. The quantity $\nu F_{\nu}$ is in arbitrary units.}}
  \label{fig:rin}
\end{figure}
\begin{figure}[hbt!]

\centerline{\epsfig{figure=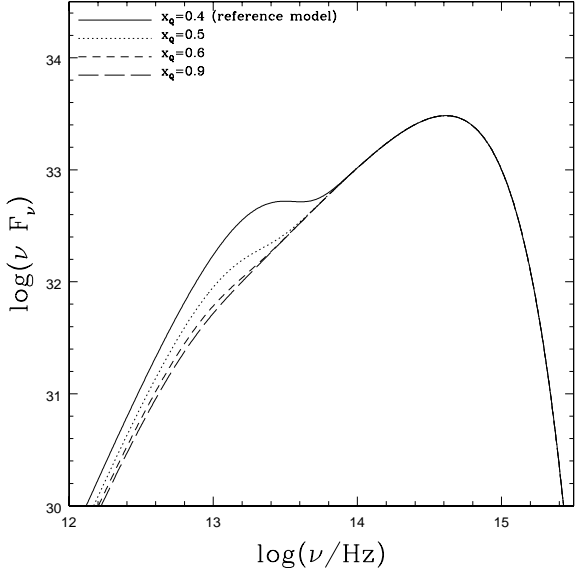,height=8cm,width=8cm}}
\caption{\small{Shape of the spectral energy distribution for different values of
the transition dimensionless radius $x_Q$. The quantity $\nu F_{\nu}$ is in 
arbitrary units.}}
  \label{fig:xq}
\end{figure}

Figure \ref{fig:rin} shows the effects of changing the inner radius of the 
disk. Not surprisingly, this parameter influences only the optical part of the 
spectrum. In Fig. \ref{fig:xq} we increase the value of $x_Q$. For self-gravity to 
give a substantial contribution to the spectral energy distribution it is necessary 
that a significant portion of the disk be self-gravitating. For example, for $x_Q=
0.6$, which corresponds to $r_Q\approx 8.5$AU (recall that the outer radius is 
$r_{out}=10$AU), the SED is still very close to that of the ``Keplerian'' case.

\subsection{Scaling relations with respect to $\alpha$} 
\label{scaling}

In the previous subsection we have described the changes in the resulting spectral 
energy distribution in response to changes in the physical parameters of the disk.
Here, we will briefly discuss the constraints posed on the relevant parameters, when
we consider the problem of fitting the SED of a given object by the self-gravitating
accretion disk model.

The highest temperature of the disk, which defines the typical frequency of the
optical-near infrared ``bump'' of the SED, is determined, as in standard accretion 
disks, by the mass accretion rate, through the relation:
\be
\label{constraint}
\sigma_BT_{in}^4\approx\frac{3G\dot{M}M_{\star}}{8\pi r_{in}^3},
\ee
independent of the value of $\alpha$. In a similar way, the typical luminosity of 
the optical-near infrared part of the SED should be given by the standard 
non-self-gravitating expression:
\be
\label{constraint2}
L_{acc}\approx \frac{G\dot{M}M_{\star}}{2r_{in}}.
\ee
Therefore, if the SED of a given object (say, an FU Orionis object) is to be 
reproduced by a simple disk model, the observed typical wavelength and luminosity of
the optical part of the SED (together with an estimate of $M_{\star}$), determines 
the value of \mdot and $r_{in}$ (the latter quantity turns out to be of the order
of a few stellar radii; Hartmann \& Kenyon \cite{hart}).
On the other hand, in the self-gravitating case, the outer disk temperature, which 
defines the typical frequency of the long wavelength excess, is expected to be 
associated with a temperature scale $T_{out}^4\approx T_0^4/\alpha$ (see Eq. 
(\ref{surftemp})). In this way, the simultaneous constraints of fitting the optical 
and the far infrared parts of the SED define both $\alpha$ and \mdot. 

Conversely, for objects such as T Tauri stars, for which the optical SED is 
dominated by the central star (see Sect. \ref{irradiation} below), the 
constraints set by Eqs. (\ref{constraint}) and (\ref{constraint2}) do not apply. In 
turn, under the hypothesis that the far infrared excess is due to the 
self-gravitating disk (which, however, may not be the case, because irradiation may 
often dominate), the temperature $T_{out}^4\approx T_0^4/\alpha$ is still 
constrained by the observations along with the luminosity scale of the far infrared 
spectrum, which is set by $L_D$. In this case, when no further information on the 
accretion rate is given, {\it for a given observed object}, from Eq. (\ref{rs}), 
Eq. (\ref{eq:deft0}), and Eq. (\ref{eq:defld}) we find the following scaling 
relations with respect to the viscosity parameter $\alpha$:

\begin{equation}
\label{scalemdot}
\dot{M}\propto \alpha,
\end{equation}
\begin{equation}
r_s\propto (\dot{M}/\alpha)^{-2/3}=const.
\end{equation}
\begin{equation}
\label{scalecostheta}
\cos\theta\propto\dot{M}^{-1}(\dot{M}/\alpha)^{-2/3}\propto\alpha^{-1}.
\end{equation}

Indeed, in Section \ref{ttau} we will see that, for two of the T Tauri objects 
considered, the exact value of $\alpha$ is not constrained solely by the observed 
far infrared excess, so that there is room for using different values of $\alpha$. 
In the other cases shown, the disk gives a contribution also to the higher frequency
part of the SED, and so there is little leverage on $\alpha$, as in the disk 
dominated case.

\subsection{Inclusion of the central star and of disk irradiation}
\label{irradiation}

During FU Orionis outbursts the luminosity of the disk is much higher than that of
the central star, so this latter contribution is generally negligible. In contrast, 
in the case of T Tauri objects the situation is reversed, so that the stellar 
luminosity is much higher than that of the accretion disk. There can be however 
mixed situations in which both contributions should be taken into account. In this 
case, we should add to the right hand side of Eq. (\ref{diskflux}) the contribution 
from the central star, which, for the limited purpose of this paper, can be taken to
be that of a blackbody with given temperature:
\begin{equation}
\nu F_{\nu\star}=\frac{\pi R_{\star}^2}{D_0^2}\frac{4\pi\hbar\nu^4}{c^2}\frac{1}
{e^{h\nu/kT_{\star}}-1}~.
\end{equation}
For convenience, we define a frequency $\nu_{\star}=kT_{\star}/2\pi\hbar$. The 
resulting spectrum in terms of the dimensionless quantities defined in Sect.
\ref{spectral} is then given by:
\begin{equation}
4\pi D_0^2\nu F_{\nu}=\frac{15L_{\star}}{\pi^4}\left(\frac{\nu_0}{\nu_{\star}}
\right)^4\frac{\hat{\nu}^4}{e^{\nu_0\hat{\nu}/\nu_{\star}}-1}+
L_D\hat{\nu}^4\int_{x_{in}}^{x_{out}}\frac{x\de x}
{e^{\hat{\nu}/\hat{T}(x)}-1},
\end{equation}
where:
\begin{equation} 
L_{\star}=4\pi R_{\star}^2\sigma_B T_{\star}^4.
\end{equation}

In addition to its direct contribution, the presence of the star affects the 
spectral energy distribution also through the reprocessing of the starlight by the
disk. For a flat disk, the integrated effect of this contribution can be as high as 
$L_{\star}/4$. The effects of disk irradiation can be calculated by following 
Adams, Lada \& Shu (\cite{shu}), as an additional contribution to the heat balance 
equation, so that a term $F_D(r)$ should be added to the right-hand side of Eq. 
(\ref{surftemp}), with:
\begin{eqnarray}
\nonumber F_D(r)= & \displaystyle\frac{\sigma_BT_{\star}^4}{\pi} & 
\left[\arcsin\left(\frac{R_{\star}}{r}\right)-\frac{R_{\star}}{r}
\sqrt{1-\left(\frac{R_{\star}}{r}\right)^2}\right]\simeq\\
 &\displaystyle\frac{2\sigma_BT_{\star}^4}{3\pi} & \left(\frac{R_{\star}}{r}
\right)^3,
\end{eqnarray}
for a flat disk, and where the approximation is valid for $r\gg R_{\star}$. 
In terms of the dimensionless quantities defined previously, we find:
\begin{equation}
\label{def:a}
\frac{F_D}{\sigma_B T_0^4}=\frac{2}{3\pi}\left(\frac{T_{\star}}{T_0}\right)^4
\left(\frac{R_{\star}}{r_s}\right)^3\frac{1}{x^3}=\frac{a}{x^3}.
\end{equation} 
The first term on the right-hand-side of Eq. (\ref{surftemp}), associated with
viscous dissipation, gives a dimensionless contribution to the surface temperature
profile $\simeq 1/x^3$, so that the parameter $a$ defined in Eq. (\ref{def:a}) 
measures the relative importance of irradiation versus viscous dissipation. On the 
other hand, at large radii, when self-regulation becomes important, the leading term
that determines the surface temperature is the second, Jeans related term, on the
right-hand-side of Eq. (\ref{surftemp}). It can be easily seen that 
\begin{equation}
a=\frac{2}{3\pi}\left(\frac{R_{\star}}{GM_{\star}}\right)\frac{L_{\star}}
{\dot{M}},
\end{equation}
where $R_{\star}$, $M_{\star}$, and $L_{\star}$ are the stellar radius, mass, and 
luminosity, respectively. As expected, the relative importance of the stellar 
irradiation grows with increasing stellar luminosity and decreases with increasing 
mass accretion rate.

In the following, we will generally neglect this irradiation term. We will check
{\it a posteriori the consistency of this assumption} in the fitting procedure in 
the case of T Tauri stars, by calculating $a$ on the basis of the fitted parameters.

In the case of disk flaring, it has been shown (Ruden \& Pollack \cite{ruden}) that 
the effect of irradiation is enhanced as:
\begin{equation}
F_D(r)=\frac{2\sigma_BT_{\star}^4}{3\pi}\left(\frac{R_{\star}}{r}
\right)^3\left[1+\frac{1}{2}\left(\frac{h}{R_{\star}}\right)\left(\frac
{\de\ln h}{\de\ln r}-1\right)\right],
\end{equation}
in the limit $h,R_{\star}\ll r$. The term $({\de\ln h}/{\de\ln r}-1)$ is determined 
from the detailed vertical structure of the disk; its value is 
equal to $1/9$ in the model by Kenyon \& Hartmann (\cite{kenyon2}) or to $2/7$ in 
the model by Chiang \& Goldreich (\cite{chiang}). In our model (see BL99) at large 
radii the opening angle $h/r$ does not depend on $r$ and hence there is no flaring.

\subsection{Simple estimates of \mdot and \mdisk}
\label{simple} 
Here we will briefly discuss the simple arguments often provided (see Kenyon \& 
Hartmann \cite{kenyon2}) to estimate the values of the accretion rate \mdot and of
the disk mass \mdisk directly from the observations of T Tauri stars.

{\it In the case in which viscous dissipation is the only source of heating}, the 
disk surface temperature can be expressed approximately as:
\begin{equation}
\sigma_BT_s^4(r)=-r\Omega\frac{\dot{M}}{4\pi}\frac{\de \Omega}{\de r}= 
\frac{\dot{M}\Omega^2}{4\pi}\left|\frac{\de\ln\Omega}{\de\ln r}\right|\approx
\frac{G\dot{M}M_{tot}(r)}{4\pi r^3},
\label{eq:simple}
\end{equation}
where, in the last expression, $M_{tot}=M_{\star}+M_{disk}$. This expression, apart 
from numerical factors of order unity, is valid both in the Keplerian case (where 
$M_{tot}\approx M_{\star}$ and $\Omega^2\approx GM_{\star}/r^3$) and in the disk
dominated ``flat rotation curve'' case, where $M_{disk}(r)\propto r$ and $\Omega
\propto 1/r$. For a disk with a transition from an inner Keplerian region to an 
outer self-gravitating region with flat rotation curve, the mass of the disk in its 
outer part is given by $M_{disk}(r)/M_{\star}\approx r/r_s$, where $r_s$ is the 
transition radius (see Eq. (\ref{rs})).

If we try to follow the argument by Kenyon \& Hartmann (\cite{kenyon2}), for the 
case of T Tauri stars, in order for the outer disk to produce the required infrared 
excess at wavelengths of the order of $25-100\mu$m, its temperature should be of the
order of 100K. Therefore, the required product $\dot{M}M_{tot}(r_{out})$ can be 
estimated by fixing $T_s(r_{out})=100$K (with $r_{out}\approx 20$AU; but note that a
value of $r_{out}\approx 10$AU would be more appropriate if we refer to the models 
described in Sect. \ref{ttau} below). With this procedure one gets $\dot{M}M_{tot}/
(10^{-7}M_{\odot}^2yr^{-1})\approx 10^3$. Kenyon \& Hartmann (\cite{kenyon2}) then 
noted that both alternatives left by an interpretation in terms of an active disk 
would be highly implausible: ({\it i}) the accretion rate is high, at least in the 
outer disk, while the disk mass remains very small, thus leading to far too high 
values of the accretion rates ($\dot{M}>10^{-5}M_{\odot}/yr$), or ({\it ii}) the 
disk mass at $r_{out}$ is very high, of the order of $1000M_{\odot}$, while the 
accretion rate remains small ($\dot{M}\lesssim 10^{-7}M_{\odot}/yr$). In this latter
case, to have such a high disk mass, the transition radius to the self-gravitating 
part of the disk should be very small. In fact, according to Kenyon \& Hartmann 
(\cite{kenyon2}) the transition would occur at $r=R_{\star}$. 

In reality, in a truly self-gravitating case, {\it viscous dissipation is not the
only source of heating} (see discussion in Sect. \ref{energy} and BL01), and so Eq. 
(\ref{eq:simple}) cannot be used. In addition, when a detailed model of 
self-gravitating accretion disk of the kind adopted in this paper is considered, on
sees that the transition radius is not an independent parameter. In particular, from
Eq. (\ref{rs}), when $M_{\star}=0.6M_{\odot}$ and $\dot{M}=10^{-7}M_{\odot}/yr$, we 
see that $r_s\approx 40$AU, so that the implied disk mass need not to be so high. On
the other hand, when self-gravity effects are fully incorporated, the surface 
temperature is going to be higher than expected from the simple estimates associated
with Eq. (\ref{eq:simple}). In fact, from Eq. (\ref{surftemp}), especially when 
$\alpha$ is small, it is clear that even with mass accretion rates lower than those 
inferred from arguments based on Eq. (\ref{eq:simple}) for the ``small-disk-mass'' 
active disk, an active disk could produce the required luminosity. 

In Section \ref{ttau} this issue will be discussed further, in view of the 
parameters determined by fitting concrete examples of spectral energy distributions
of T Tauri stars. Here we would like to anticipate that, even though the results, as
already suggested in this subsection, show that the required mass accretion rates 
need not be as high as estimated earlier, for the majority of T Tauri stars, the 
picture of a pure, self-gravitating, active disk remains unsatisfactory.

\section{A possible fit to the spectral energy distribution of some FU Orionis 
systems}
\label{fit}

In this Section we will compare the model described in the previous Sections with
the observations by fitting the spectral energy distribution of some FU Orionis 
objects. In these systems there is already evidence that the accretion disk is {\it 
active}, i.e. that most of the emission from these objects is probably due to the 
disk itself, which outshines the central star. There are already clues that disk 
self-gravity should play a major role in these cases; in particular, most outburst 
models predict the existence of massive disks (Bell et al. \cite{bell}; Hartmann \& 
Kenyon \cite{hart}).

\subsection{A sample of FU Orionis systems}

\begin{table}[hbt!]
\begin{center}
\begin{tabular}{cccc}
\hline
\hline
star      & $M_{\star}/M_{\odot}$ & $A_V$ & $D_0$   \\
\hline
\object{FU Ori}    &      1                &   2   & 550  pc \\
\object{V1515 Cyg} &      1                &   2.8 & 1000 pc \\

\hline

\end{tabular}
\end{center}
\caption{\small{Stellar parameters assumed for the FU Orionis sample stars.}}
\label{tab:fuor}
\end{table}

Currently, there is a handful of known FU Orionis objects, among which the three
best studied systems are FU Ori, V1515 Cyg, and \object{V1057 Cyg}. The last object 
shows a relatively rapid decline in B magnitude after the outburst, with a decay 
timescale of a few years, about ten times faster than FU Ori, and thus special care 
must be taken when comparing observations made at different epochs. For this reason 
we have not included V1057 Cyg in our sample of FU Orionis systems. In addition to 
FU Ori and V1515 Cyg, we have also studied V1735 Cyg, which is an ``extreme'' case, 
with a huge infrared excess, so that the infrared luminosity is apparently larger 
than the optical one. Actually, this object is heavily obscured, so that it is very 
difficult to derive the extinction correction with confidence and hence to extract 
for it the intrinsic spectral energy distribution. Therefore, although we initially 
kept this object in our sample, so as to have an idea of how large the disk mass 
should be when the infrared excess is so large, we do not record here the results of
the fit for this case.

The parameters to be determined by the fit are: the characteristic frequency
$\nu_0$, the dimensionless inner and outer radii of the disk $x_{in}$ and $x_{out}$,
the inclination angle $\theta$, and the viscosity parameter $\alpha$. We have 
derived the best-fit parameters for different choices of $x_Q$ taken to vary in the 
interval $[0.4,0.9]$. To derive the physical parameters of the disk (such as the 
accretion rate and the disk mass), we have to specify the central star mass (see 
Table \ref{tab:fuor}). In the case of V1735 Cyg, we have made a fit (not shown in 
detail here) by assuming $A_V=0,~M_{\star}=2M_{\odot}$, on the high side, given the 
unusual infrared luminosity of the object (see also Sect. \ref{spectral}). In 
any case, in contrast with the models that are generally considered, we do not 
include here irradiation by the inner on the outer disk (our outer disk has no 
flaring and hence this contribution is negligible) and the possible contribution of 
an additional infalling envelope.

\begin{table*}[hbt!]
\begin{center}
\begin{tabular}{cccc}
\hline
\hline
                       & FU Ori                        & V1515 Cyg   \\  
\hline
$\mbox{Log}(\nu_0/Hz)$ & [12.00-12.09]                 & [11.94-12.12] \\
$\cos\theta$           & [0.65-0.72]                   & [0.94-0.88]   \\
$\alpha$               & [$3.4~10^{-2}$-$1.61~10^{-2}$]& [$2.17~10^{-3}$-
$1.09~10^{-3}$] \\
$x_{in}$               & [0.0013-0.0018]               & [$9~ 10^{-4}$-0.0016]  \\
$x_{out}$              & [1.52-2.51]                   & [1.04-1.96]   \\

\hline

\end{tabular}
\end{center}
\caption{\small{Best-fit parameters for two FU Orionis objects. 
The values in square brackets refer to different choices of the 
free parameter $x_Q$ ranging in the interval $[0.4-0.9]$.}}
\label{tab:fuorresults}
\end{table*}
\begin{table*}[hbt!]
\begin{center}
\begin{tabular}{cccc}
\hline
\hline
                                & FU Ori                        & 
V1515 Cyg                      \\ 
\hline
$\dot{M}/(10^{-5}M_{\odot}/$yr) & [8.2-6.6]                     & 
[1.1-1.2]                       \\
$M_{disk}/M_{\odot}$            & [1.7-2]                       & 
[0.91-1.52]                     \\
$r_s/$AU                        & [38-26]                       & 
[23-14]                         \\
$\theta$                        & [$49^{\circ}-44^{\circ}$]     & 
[$20^{\circ}-28^{\circ}$]       \\

\hline

\end{tabular}
\end{center}
\caption{\small{Derived parameters for the FU Orionis objects. The brackets refer to
the same choice of $x_Q$ as in Table \ref{tab:fuorresults}}}
\label{tab:fuorresults2}
\end{table*}

\begin{figure*}[hbt!]
  \centerline{\epsfig{figure=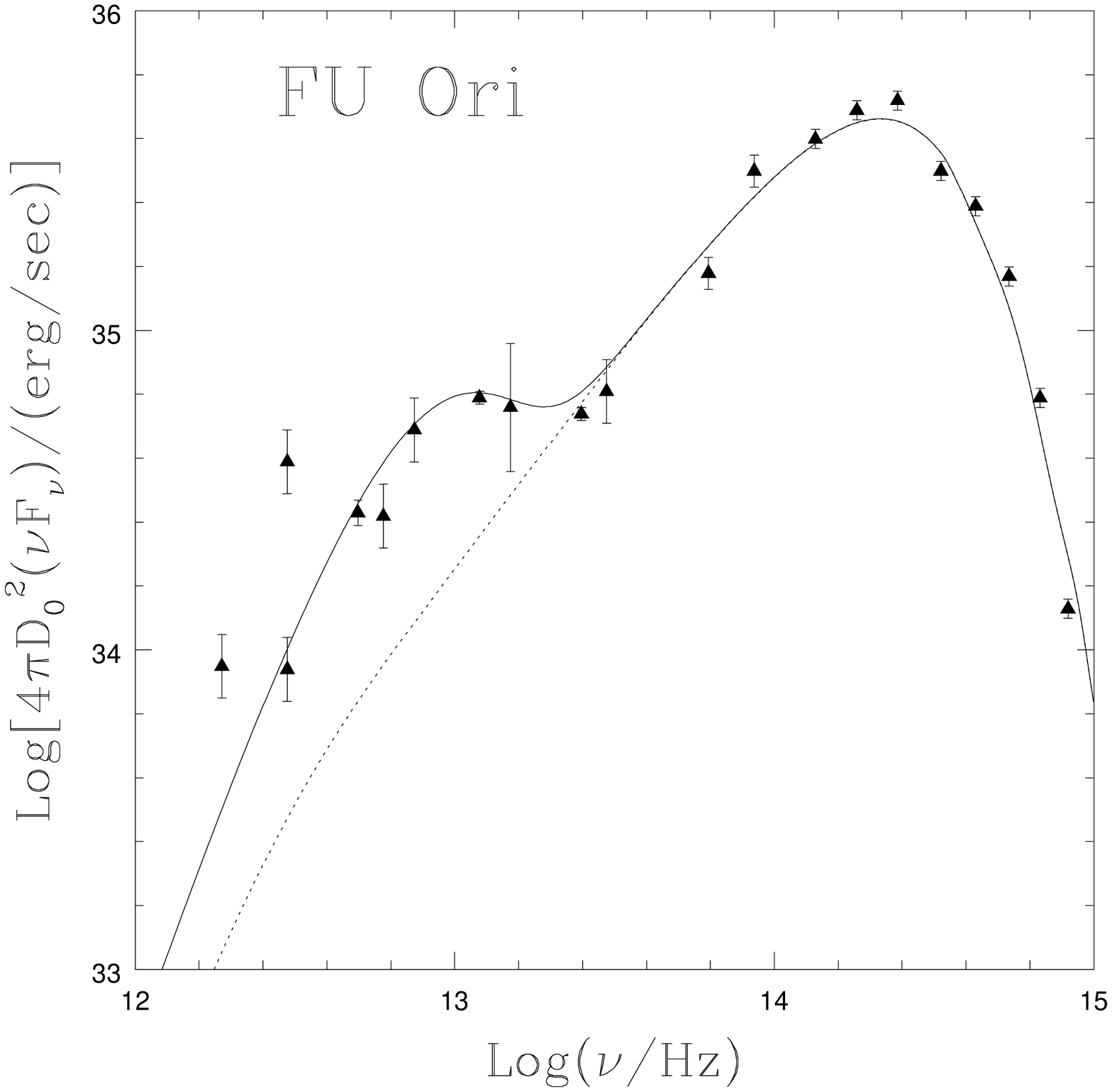,height=7.5cm,width=7.5cm}
	      \epsfig{figure=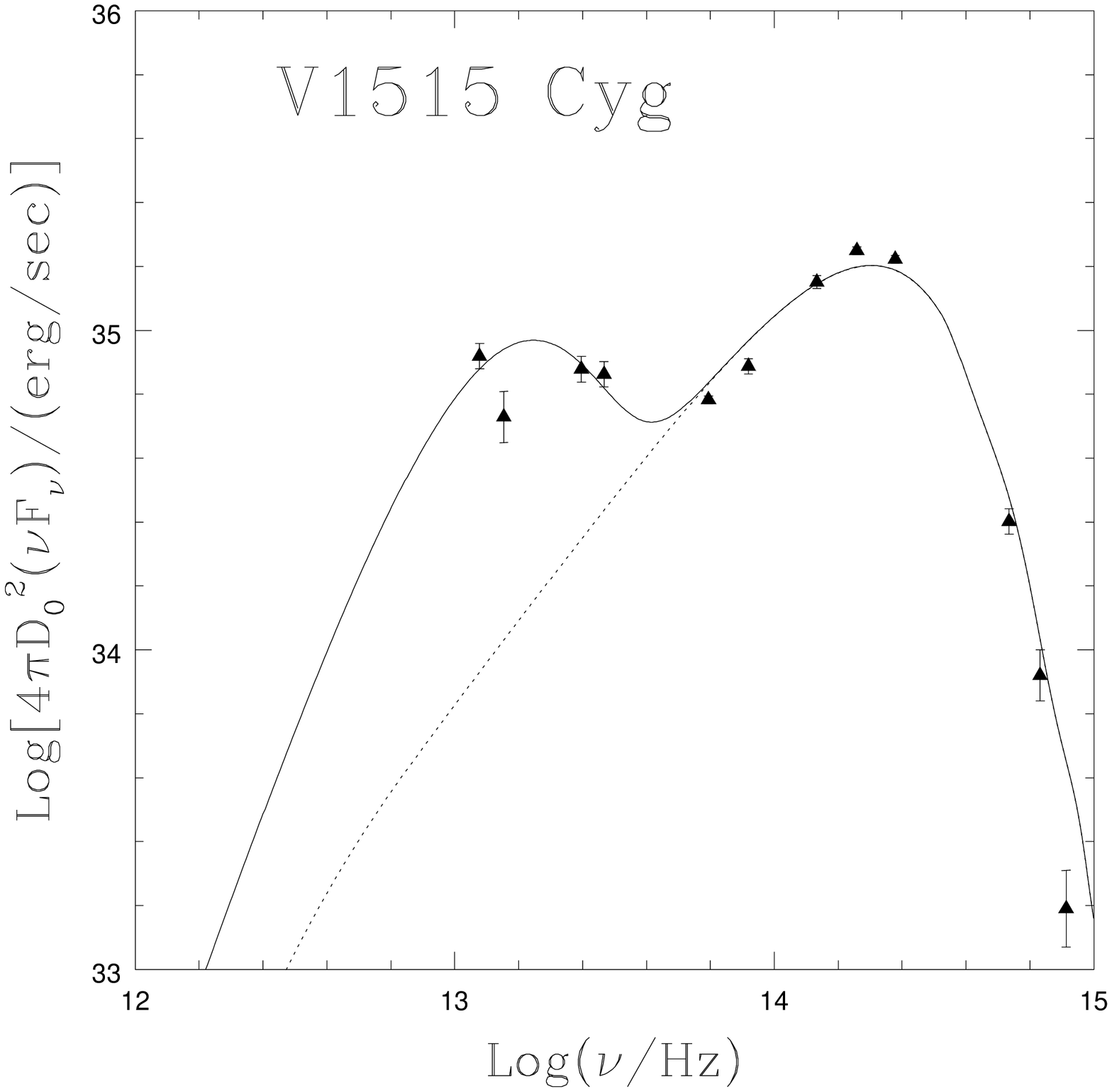,height=7.5cm,width=7.5cm}}

\caption{\small{Best fit to the spectral energy distributions for the FU Orionis 
sample (solid curves) compared to the spectra resulting from Keplerian disks with 
the same parameters ($\dot{M}$, $\cos\theta$, $x_{in}$, $x_{out}$, $\alpha$; thin 
curves). Data from Kenyon \& Hartmann (\cite{kenyon3}) and from 
Weaver \& Jones (\cite{weaver}).}}
  \label{fig:fuor}
\end{figure*}

The optical and near-infrared luminosities are those collected by Kenyon \& 
Hartmann (\cite{kenyon3}). The IRAS fluxes are taken from Weaver \& 
Jones (\cite{weaver}). First the data have been dereddened using the procedure of 
Cardelli, Clayton \& Mathis (\cite{cardelli}) and then the spectral energy 
distributions described in the previous Section have been fitted to the data. 

\begin{figure*}[hbt!]
\centerline{\epsfig{figure=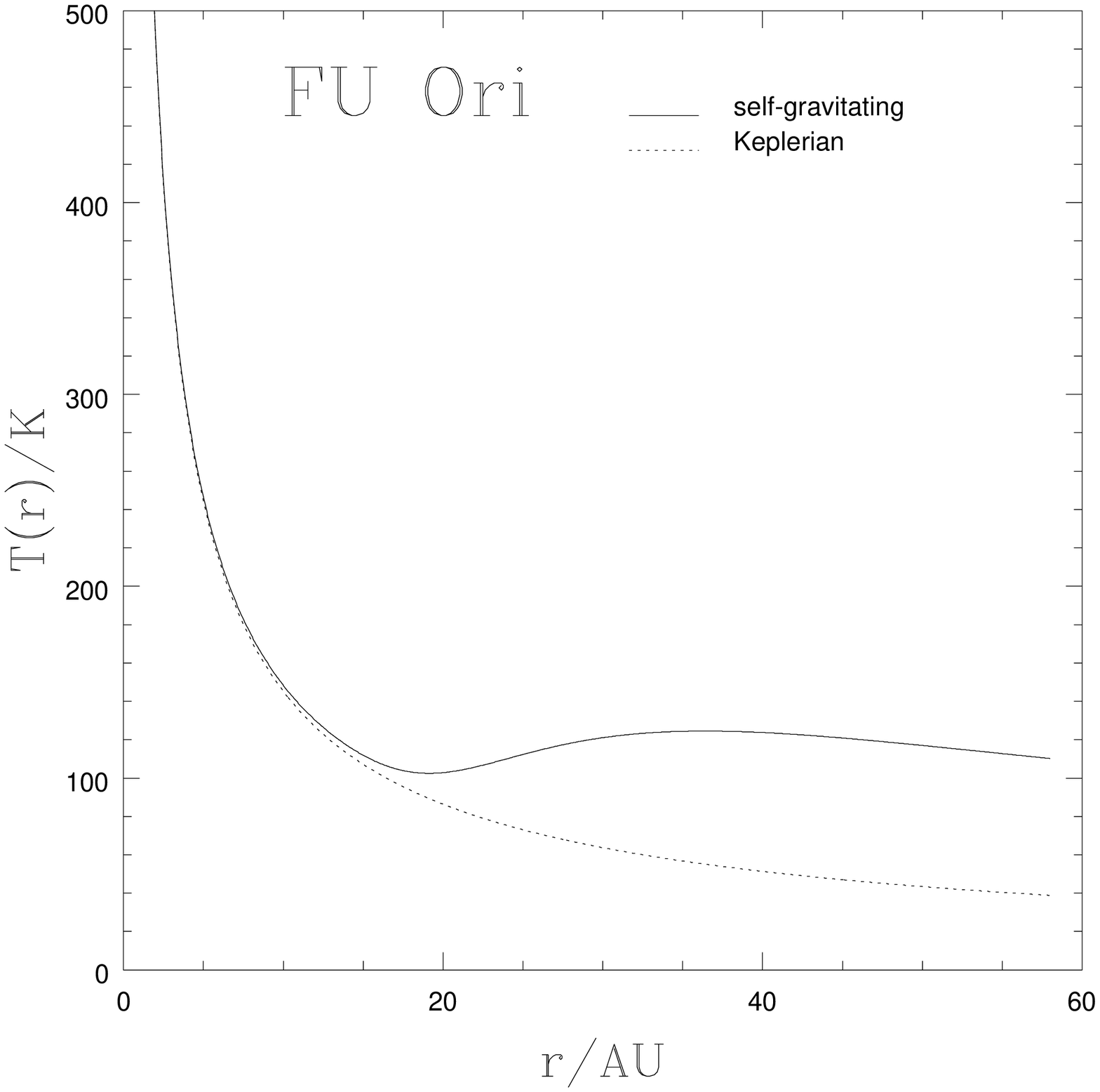,height=7.5cm,width=7.5cm}
	    \epsfig{figure=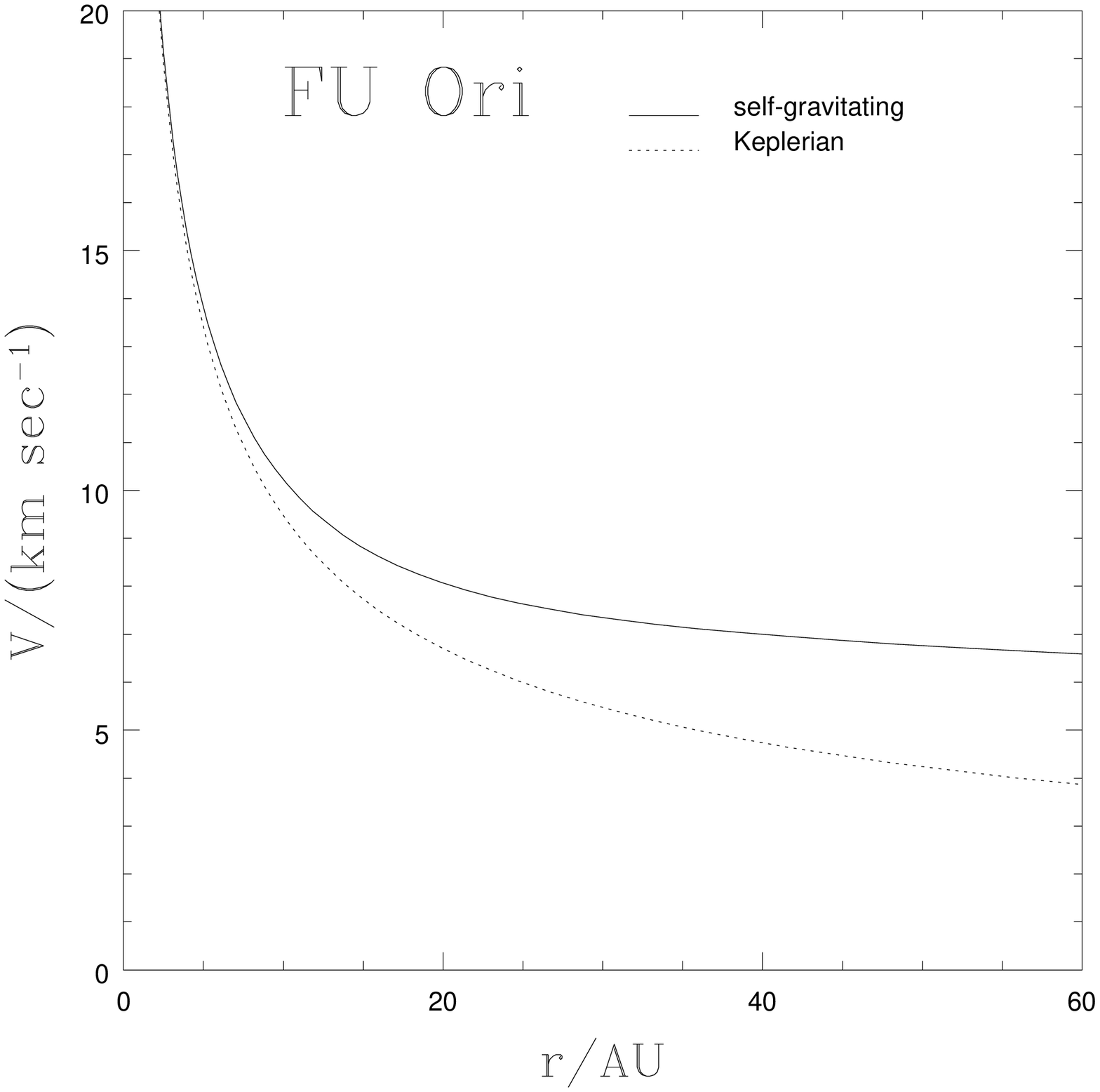,height=7.5cm,width=7.5cm}}
\caption{\small{Surface temperature profile (left) and rotation curve (right) for 
the best-fit model to FU Ori (with $x_Q=0.4$), compared with the corresponding 
Keplerian model.}}
  \label{tempproffu}
\end{figure*}

The best-fit parameters are shown in Table \ref{tab:fuorresults} and the derived 
physical parameters of the disk are shown in Table \ref{tab:fuorresults2}. Figure
\ref{fig:fuor} displays the spectral data of the stars together with our best-fit 
SEDs (solid lines) and the corresponding SEDs for non-self-gravitating, Keplerian 
disks (thin lines), with surface temperature $T_s\propto r^{-3/4}$. In Fig. 
\ref{tempproffu} we plot, as an example, the surface temperature profile and 
the rotation curve for the best-fit model to FU Ori (in the case $x_Q=0.4$) together
with the corresponding quantities for the Keplerian model, showing that the outer 
parts of the disk in the self-gravitating case are indeed hotter and rotate faster 
than the corresponding Keplerian disk.

The mass accretion rates are of the order of $10^{-5}-10^{-4}M_{\odot}/$yr. The 
value of $\alpha$ {\it determined by the fit} is in the range $10^{-3}-10^{-2}$, 
consistent with some theoretical expectations. Typical disk radii are of the order
of $50$AU. Disk masses derived from the fit are of the order of one solar mass. Some
of these conclusions would also apply to V1735 Cyg, except that in this case, as a 
result of its very high infrared excess, the estimated disk mass would be around 
$20M_{\odot}$. 

\subsection{Discussion}

The infrared spectral energy distribution of FU Orionis systems has been described
in terms of active accretion disks by Kenyon et al. (\cite{hewett}) and by Kenyon \&
Hartmann (\cite{kenyon3}). These authors find good agreement between the accretion 
disk model and the observed spectra for wavelengths smaller than $10\mu$m. The 
excess emission at larger wavelengths is attributed by Kenyon \& Hartmann 
(\cite{kenyon3}) to a flattened envelope which reflects the luminosity of the disk. 
The hypothesis that this excess comes from a flared, outer disk is discarded by 
Kenyon \& Hartmann (\cite{kenyon3}), because it would have required unreasonable 
levels of flaring.

Our estimates of the mass accretion rates are very close to those by Kenyon \& 
Hartmann (\cite{kenyon3}), because these numbers are determined mainly by the 
optical-near infrared part of the spectral energy distribution (see also Sect. 
\ref{scaling}); indeed, as can be seen from Fig. \ref{fig:fuor}, the spectrum for 
$\lambda<10\mu$m is not affected by self-gravity related effects. On the other hand,
in contrast with the interpretation of Kenyon \& Hartmann (\cite{kenyon3}), we 
attribute in this paper the far infrared emission to the self-gravitating part of 
the disk. The mass thus required to produce the observed excess is of the order of 
one solar mass, which is rather high, but in line with the indications of current 
outburst models (Bell et al. \cite{bell}). Unfortunately, no separate measurements 
of the disk mass in FU Orionis systems is available to our knowledge.

The interpretation in terms of the presence of an infalling envelope responsible for
the far infrared excess is actually appealing because it can also offer a clue to 
the peculiar time evolution of V1057 Cyg, which shows a correlation between
the decay timescale in the optical and that at longer wavelengths ($10\mu$m). Here 
we do not address the time evolution of our model, and indeed we have preferred to 
leave V1057 Cyg out of our sample. Still, it would be interesting to consider these 
issues in the context of self-gravitating disks (see also the outburst model of 
Armitage et al. \cite{armitage}, which includes some effects related to the disk
self-gravity). The far infrared SED in our models is sensitive to the relative 
positions of the disk outer radius $r_{out}$ and the radius $r_Q$ at which the 
disk becomes self-regulated (see Sect. \ref{parameters}). A detailed 
description of the role of self-gravity in the context of time-dependent outburst 
models, which might result from the interplay between $r_Q$ and $r_{out}$ as a
function of time, is, however, beyond the goals of this paper.

\section{Difficulties of the self-gravitating scenario in the context of T Tauri 
stars}
\label{ttau}

In the context of T Tauri stars the situation seems to be less favorable for disk
self-gravity to play a major role, because the infrared excess is likely to result 
in general from irradiation of the disk from the central star. Many points are often
mentioned in favor of this interpretation, among which we may recall the explanation
of the observed silicate emission features at $10\mu$m (Calvet et al. 
\cite{calvet2}) and other HST observations (e.g., see Stapelfeldt et al. 
\cite{stapelfeldt}). On the other hand, initial efforts at explaining the SEDs of T 
Tauri stars with active accretion disks led to estimated mass accretion rates so 
high that the corresponding optical-infrared emission of the disk should outshine 
the central star, which is not observed. In addition, the peculiar ``flat'' shape of
the far infrared SED, when explained in terms of a non-Keplerian disk, led to 
unreasonably high disk masses (Kenyon \& Hartmann \cite{kenyon2}; Shu, Adams \& 
Lizano \cite{lizano}). 

The model described in this paper presents significant, qualitative and 
quantitative, differences with respect to the non-Keplerian models considered in 
previous investigations. First, the additional heating term in Eq. (\ref{surftemp}) 
gives an important contribution to the heating of the disk in its self-gravitating 
part, so that the corresponding mass accretion rates needed to account for the 
infrared luminosity in an active disk scenario may be lower than previously 
estimated. Second, as pointed out in Sect. \ref{simple} and as the fit to FU 
Orionis objects in the previous Section demonstrates, the disk masses needed to give
a significant contribution to the SED in the self-gravitating scenario are not as 
high as previously argued. Therefore, we think it appropriate to reassess the issue 
of the required parameters within the self-gravitating active disk scenario.

Furthermore, in some cases (see Gullbring et al. \cite{gullbring2}; Testi et al. 
\cite{natta}), such as in some intermediate mass Young Stellar Objects, the disk
mass may be relatively high, so that the self-gravity of the disk should play some 
role, even without abandoning the general framework of irradiated disks. In 
addition, as noted in Sect. \ref{simple}, when $M_{\star}\approx 0.6M_{\odot}$
and $\dot{M}\approx 10^{-7}M_{\odot}/$yr (which is not unreasonable at least for
some objects), the implied values of $r_s$ readily show that the contribution of
disk self-gravity should be incorporated. 

For these reasons we now proceed to fit the SEDs of some T Tauri stars with the 
model presented in this paper. The primary goal is to check in detail the parameter
requirements for the active, self-gravitating disk scenario, so as to confirm or to
correct previous estimates, even though this model is going to be relevant only for 
a small set of less typical cases.

\subsection{A sample of T Tauri stars}

We have considered two typical T Tauri stars (\object{BP Tau} and \object{DE Tau}) 
in the Taurus-Auriga molecular cloud (at a distance $D_0=140$ pc) for which the 
stellar masses are available from Gullbring et al. (\cite{gullbring}) and the 
stellar temperatures and visual extinction from Kenyon \& Hartmann (\cite{kenyon}). 
In addition, to check the parameter requirements on our models set by less typical 
cases, we have considered two ``continuum'' T Tauri stars, \object{DR Tau} and 
\object{DG Tau}, in which the high amount of veiling in the optical suggests the
existence of high accretion rates (Gullbring et al. \cite{gullbring2}). DG Tau is 
also one of the ``flat spectrum'' T Tauri stars considered by Adams et al. 
(\cite{shu}). The stellar parameters for the last two cases have been taken from 
Gullbring et al. (\cite{gullbring2}), with stellar temperature $T_{\star}=4000$K and
assumed stellar mass of $0.5M_{\odot}$. 

To fit this set of T Tauri stars, we add to the disk emission also the direct 
contribution of the star, as described in Sect. \ref{irradiation}. 
Nevertheless, we still neglect the effect of irradiation. The {\it consistency} of
this assumption will be checked {\it a posteriori} by estimating the parameter $a$ 
from the derived fit parameters. This, of course, translates into an actual 
overestimate of the mass accretion rate, as derived by the fit. The luminosity of 
the star $L_{\star}$ is added to the set of parameters to be determined by the fit. 

The optical and near-infrared luminosities for all stars are available from Kenyon 
\& Hartmann (\cite{kenyon}). The IRAS fluxes for all stars are taken from Weaver \& 
Jones (\cite{weaver}). The dereddening procedure was the same as adopted for the FU 
Orionis objects in the previous Section. For BP Tau and DE Tau, we found that the 
value of the viscosity parameter $\alpha$ does not affect significantly the quality 
of the fit, so it has been excluded from the fit parameters, and the fit has been 
performed, as separate examples, for two different values: $\alpha=10^{-3}$ and 
$\alpha=10^{-4}$ (see also comment at the end of Sect. \ref{scaling}).

\begin{figure*}[hbt!]

  \centerline{\epsfig{figure=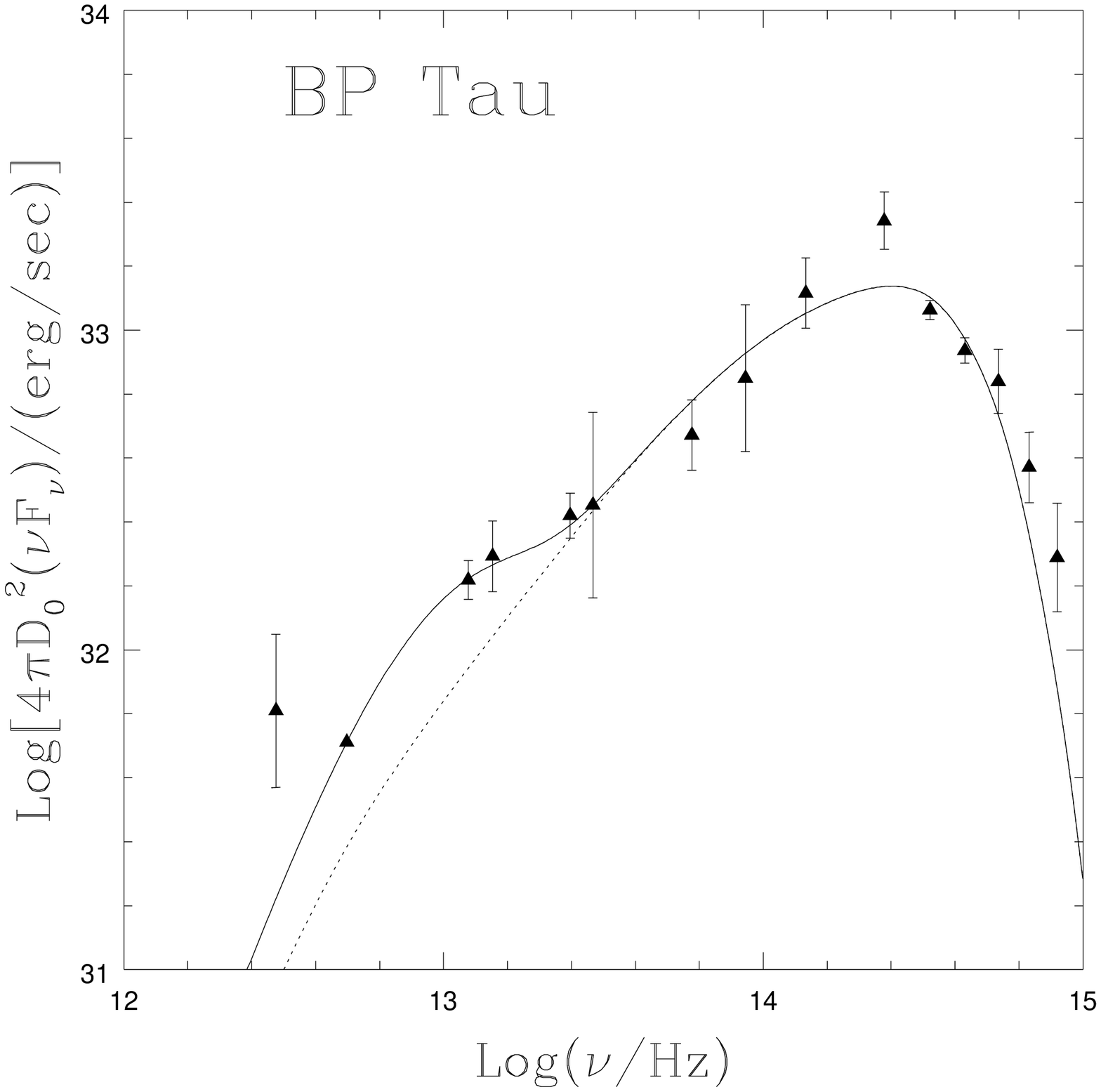,height=7.5cm,width=7.5cm}
	      \epsfig{figure=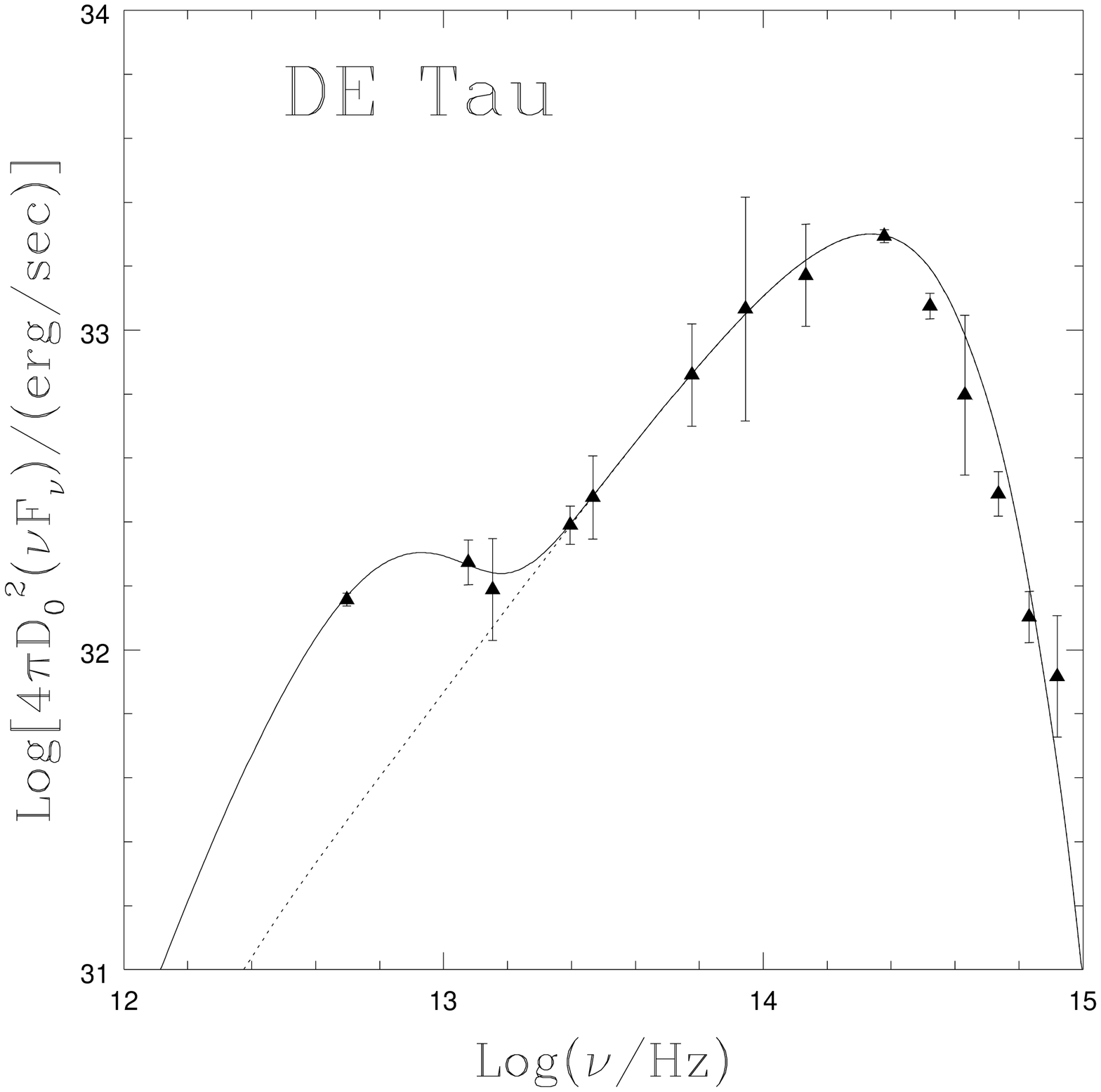,height=7.5cm,width=7.5cm}}
  \centerline{\epsfig{figure=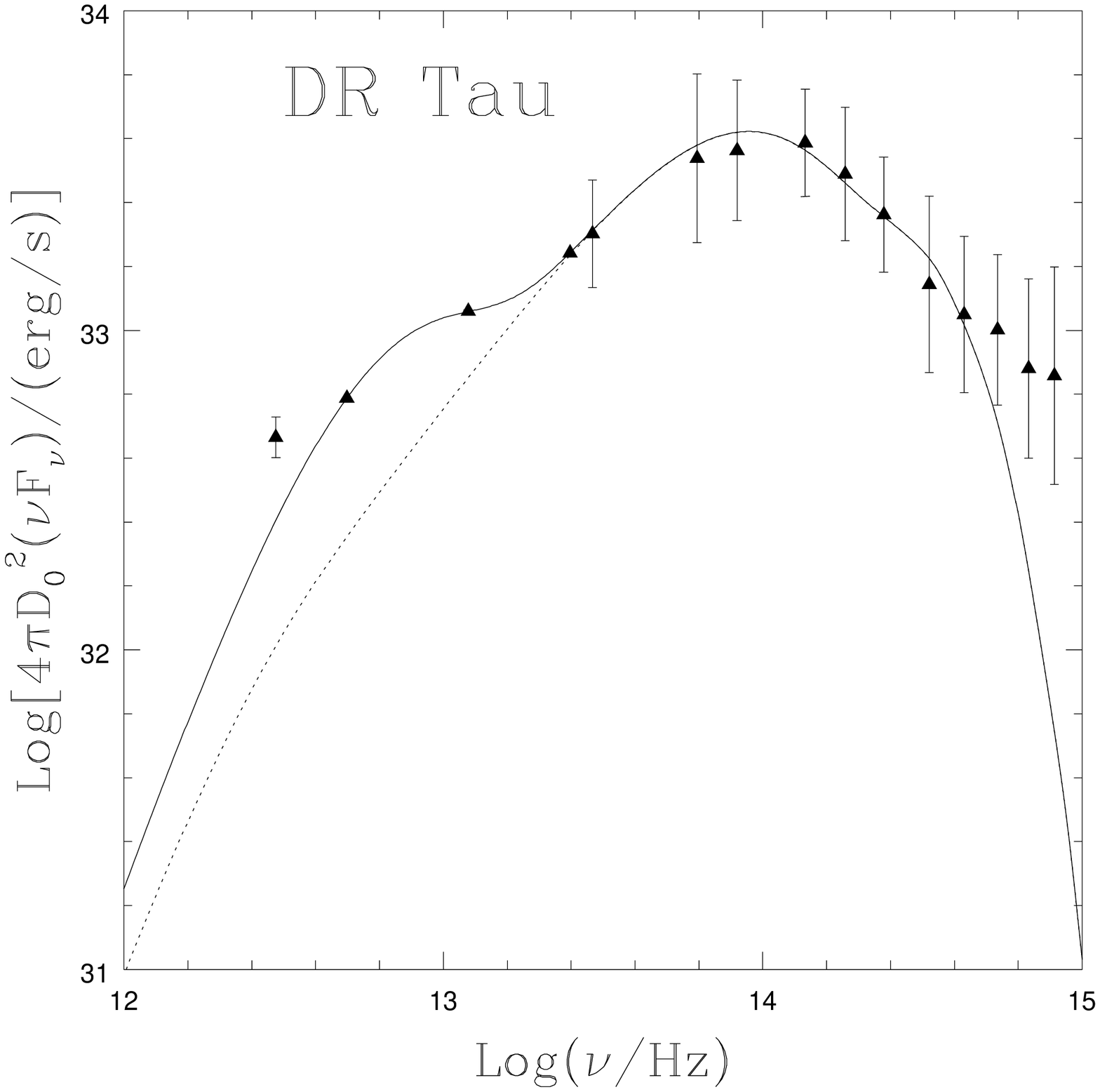,height=7.5cm,width=7.5cm}
              \epsfig{figure=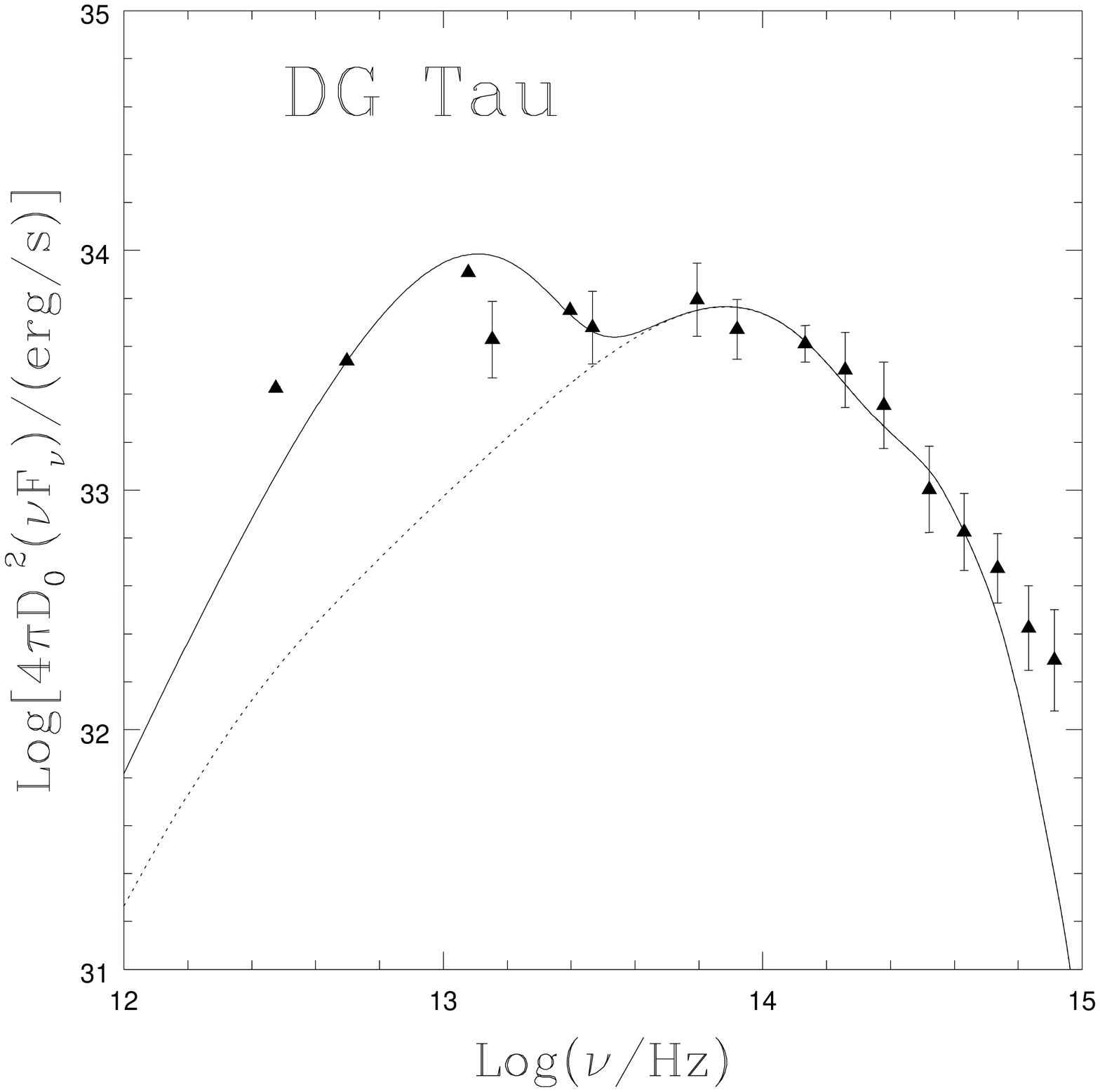,height=7.5cm,width=7.5cm}}

\caption{\small{Best fit to the spectral energy distributions of our sample stars
(solid curve), compared to the spectra resulting from Keplerian disks with the same 
parameters ($\dot{M}$, $\cos\theta$, $x_{in}$, $x_{out}$; thin curves). For all 
cases the displayed fit refers to $x_Q=0.4$. The spectral data are from Kenyon \& 
Hartmann (\cite{kenyon}) and Weaver \& Jones (\cite{weaver}).}}
  \label{fig:spectra}
\end{figure*}

\begin{figure*}[hbt!]
\centerline{\epsfig{figure=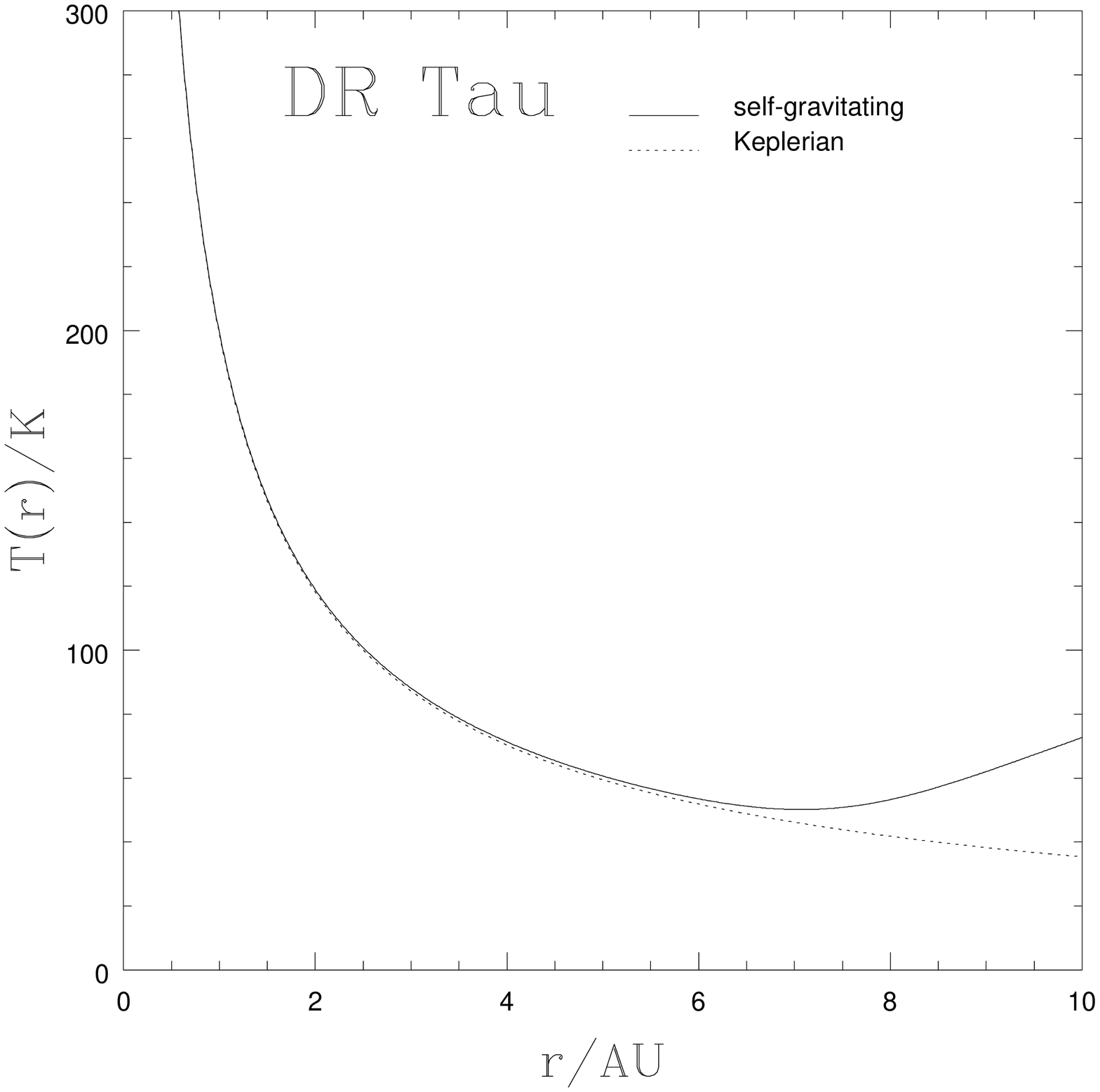,height=7.5cm,width=7.5cm}
	\epsfig{figure=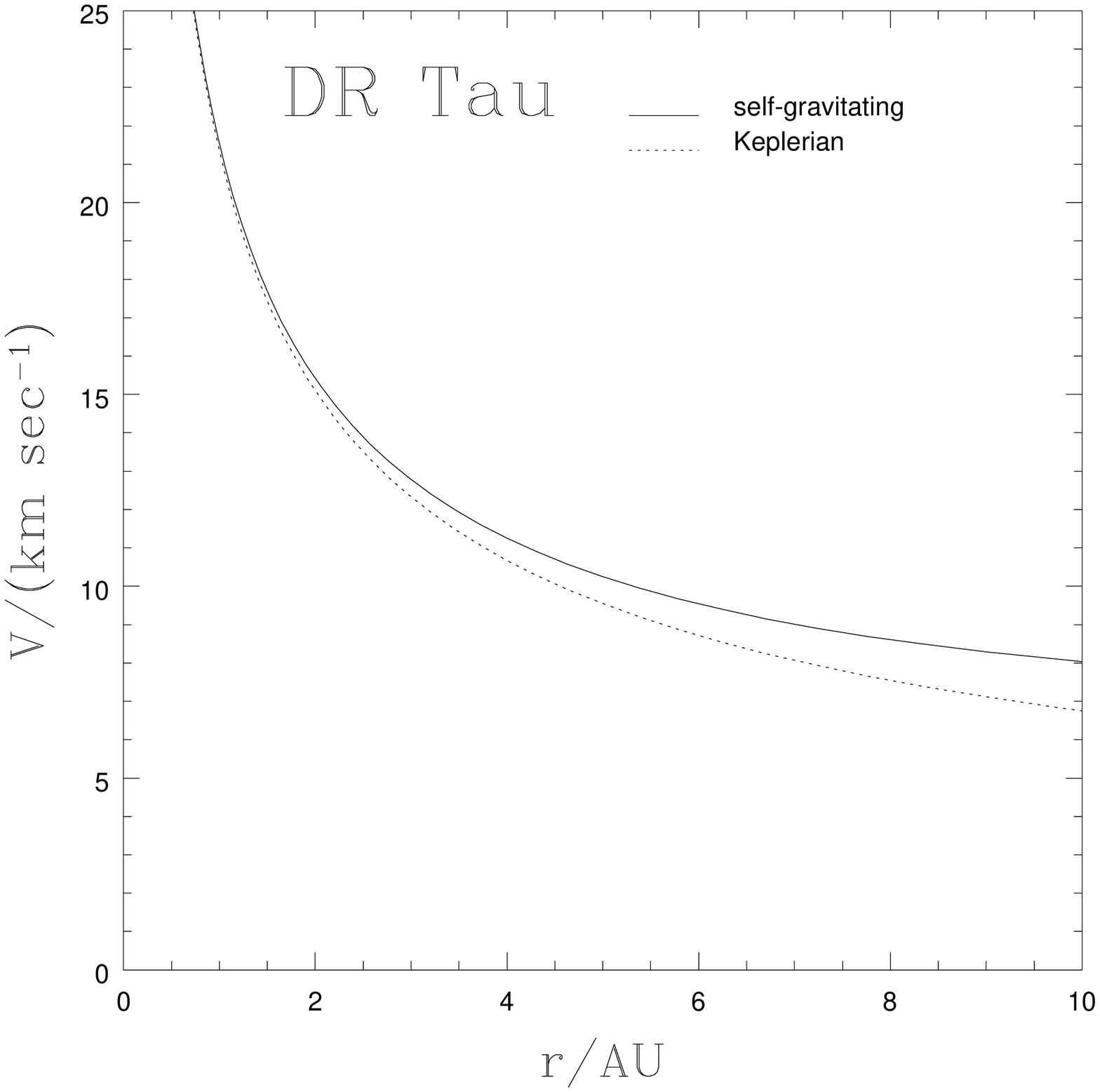,height=7.5cm,width=7.5cm}}
\caption{\small{Surface temperature profile and rotation curve for the best-fit 
model to DR Tau (with $x_Q=0.4$), compared with the corresponding Keplerian 
model.}}
  \label{tempprofttau}
\end{figure*}

The best-fit parameters (for different values of the parameter $x_Q$, taken to vary 
in the interval $[0.4,0.9]$) are shown in the Appendix in Table \ref{tab:fitpar} and
the derived parameters are reported in Table \ref{tab:derpar}. The fitted SEDs are 
shown in Fig. \ref{fig:spectra}. As can be seen, the fitted solid curves reproduce 
the available infrared data well. The surface temperature profile and the rotation
curve for DR Tau (in the case where $x_Q=0.4$) are shown in Fig. \ref{tempprofttau}.
The resulting mass accretion rates for BP Tau and 
DE Tau are of the order of $10^{-7}M_{\odot}/yr$ for $\alpha=10^{-4}$ and larger by 
a factor $\approx 10$ for $\alpha=0.001$, as anticipated from the scaling relations 
pointed out in Sect. \ref{scaling}. In these two cases, it can be seen that 
increasing the value of $x_Q$ (with $\alpha$ being fixed) leads to higher disk 
masses and accretion rates, and consequently to lower values of $r_s$ (which is 
proportional to $\dot{M}^{-2/3}$) and $a$ ($\propto\dot{M}^{-1}$). Typical disk 
radii $r_{out}$ are of the order of $10-50$ AU. The masses required to fit the data 
are of the order of some fraction of $M_{\odot}$.

For DR Tau we obtain $\dot{M}\approx 5~10^{-7}M_{\odot}/yr$, while for DG Tau we
have $\dot{M}\approx 10^{-6}M_{\odot}/yr$. The best fit value of $\alpha$ is of
the order of $10^{-4}$ and slowly decreases with increasing $x_Q$. The required 
values of the disk masses are a fraction of $M_{\odot}$ also in these cases.

\subsection{Discussion}
\label{discussion}

For T Tauri stars, the infrared spectral energy distribution is generally not 
considered to be a good tracer of the mass accretion rate, given the fact that a 
significant fraction of the infrared luminosity is likely to be due to reprocessing 
of the starlight rather than to accretion power. Mass accretion rates are better 
derived from the hot continuum expected to be produced from the accreting matter 
when it hits the stellar surface (Gullbring et al. \cite{gullbring}). In any
case, great uncertainties remain in the derived accretion rates (especially when the
hot accretion continuum is very high) mainly due to uncertainties in the extinction 
corrections. The three most extensive samples of T Tauri stars with accretion rates 
determined in this way (Valenti et al. \cite{valenti}; Hartigan et al. 
\cite{hartigan}; Gullbring et al. \cite{gullbring}) report values that differ for 
the same star by as much as a factor of ten. On the average, the Gullbring et al. 
(\cite{gullbring}) and the Valenti et al. (\cite{valenti}) estimates are consistent
with each other, although with large scatter, while the Hartigan et al. 
(\cite{hartigan}) estimates are systematically biased to higher values. 

Looking at the values in Table \ref{tab:derpar} for the cases of BP Tau and DE Tau,
it is clear that the pure self-gravitating disk scenario is unsatisfactory. For
$\alpha=10^{-3}$ the values of the required accretion rates for these objects are 
of the order of $10^{-6}M_{\odot}$/yr, two orders of magnitude larger than the 
values typically quoted (Gullbring et al. \cite{gullbring}). This is not surprising,
because it is known that an interpretation in terms of active disks requires high 
mass accretion rates. These high accretion rates also lead to unreasonably small 
inclination angles for the disks. 
One might imagine that photo-evaporation of the disk due to UV radiation field of 
the central star could reduce the mass accretion rate in the inner disk (Richling \&
Yorke \cite{richling}), thus allowing for discrepancies between the \mdot derived 
from the infrared SED (which is most sensitive to the conditions in the outer disk) 
and that obtained from the hot continuum. However, this process requires the 
presence of a strong UV source that is not available for the T Tauri stars that we
consider here. On the other hand, as suggested at the beginning of this Section, the
required disk mass is not as high as previously argued, because it turns out to be 
just a fraction of the stellar mass. 

In the case where a smaller value for $\alpha$ is assumed, the required mass 
accretion rates, in line with the scaling relation of Sect. \ref{scaling}, are 
lower, but still closer to the higher values of Hartigan et al. (\cite{hartigan}) 
than to the mutually consistent values of Valenti et al. (\cite{valenti}) and of 
Gullbring et al. (\cite{gullbring}). In this case, however, the lower accretion 
luminosity is accompanied by an increasing importance of the irradiation term, 
quantified by the higher values of $a$, thus leading to a mixed case where both 
contributions (of irradiation from the central star and of internal heating of the 
disk) are important. Finally, note that the luminosity of the star $L_{\star}$ is 
smaller than usually estimated, mainly because we attribute part of the 
near-infrared luminosity to the disk, rather than to the star.

The interplay between standard models and the self-gravitating scenario might be 
more interesting for the ``continuum'' stars, for which the best fit accretion rates
from the model studied in this paper turn out to be higher, but not dramatically, 
than what inferred from observations at shorter wavelengths (Gullbring et al. 
\cite{gullbring2}). We should reiterate that, by our modeling procedure, we
actually overestimate the mass accretion rate because the contribution of 
irradiation is neglected. 

Figure \ref{fig:spectra} shows a significant discrepancy between the model curves 
and the data at short wavelengths, particularly for DR Tau and DG Tau. Here we are
generally not concerned about the short wavelength data, because the optical and 
ultraviolet emission and its variability are well interpreted in terms of 
magnetospheric accretion (e.g., see Kenyon et al. \cite{kenyondr}). 

\section{Conclusions}
\label{conclusion}

In this paper we have described a model for self-gravitating accretion disks in the
context of protostellar disks. This model might in principle be relevant for a 
number of cases, because the mass of the protostellar disk is likely to be high in 
the earlier phases of the accretion process. In addition, massive disks are 
sometimes observed in the context of intermediate mass star formation. 

We have described the properties of the spectral energy distribution of these
self-gravitating disks and their dependence on the various parameters involved. In 
addition, we have checked the parameter requirements for self-gravitating disks to
fit the SEDs of some observed cases. In particular, we have considered the cases of
FU Orionis objects and of some T Tauri stars. 

For the FU Orionis objects we find that the infrared excess at wavelengths larger 
than $\lambda=10\mu$m can be described in terms of the presence of a 
self-gravitating disk. Our models are otherwise consistent (for the implied mass 
accretion rate, for example) with previous models (see Kenyon and Hartmann 
\cite{kenyon3}), which, in contrast, attribute the far infrared excess to an 
infalling envelope.

We have also compared the spectral energy distribution of our models with the SEDs 
of some T Tauri stars, even though in general these should not be considered for a 
realistic application of our models, in order to check the requirements on an
interpretation based on an active disk when self-gravity is taken into account. The 
required disk masses appear to be much lower than estimated previously. On the other
hand, especially when the viscosity parameter is taken to be high, we do find 
unreasonably high required mass accretion rates. For typical T Tauri stars, our very
simple models have an evident limitation, because we have not considered the 
important role of disk irradiation. 

In general, we would like to see this paper as an interesting starting point for 
the inclusion of self-gravity related effects, even in the standard framework of the
irradiated disk model.

\begin{acknowledgements}
We would like to thank L. Hartmann, for kindly providing us with the spectral data 
of FU Ori, and R. Cesaroni, A. Natta, and M. Romaniello, for interesting 
discussions. We would also like to thank an anonymous Referee, whose questions and
suggestions have led to a substantial improvement of the paper. This work has been 
partially supported by MURST of Italy.
\end{acknowledgements}

\appendix
\section{Fit parameters obtained from the SEDs of some T Tauri systems}

In this Appendix we report the detailed values of the fit parameters and of the
physical quantities derived from the fit of some T Tauri stars presented in Section
\ref{ttau}. Table \ref{tab:fitpar} shows the fit parameters (for BP Tau and DE Tau 
the parameters are shown for two adopted values of $\alpha$; see Section \ref{ttau})
and Table \ref{tab:derpar} displays the derived parameters.

\begin{table*}[hbt!]
\begin{center}
\begin{tabular}{ccccccc}
\hline
\hline
	& BP Tau & DE Tau & BP Tau & DE Tau & DR Tau & DG Tau\\
\hline
$\mbox{Log}(\nu_0/Hz)$      & [11.86-12.11] & [11.55-11.81]                     & 
[11.77-12.03]                      &  [11.43-11.68]          & [11.60-11.81]    &
[11.68-11.87]\\
$\mbox{Log}(L_{\star}/(erg/s))$   & 33.298   & 33.231        & 33.298        & 
33.231               & 33.562     & 33.499       \\
$\mbox{Log}(L_D/(erg/s))$   & [29.80-30.14]  & [29.42-29.77] & [29.67-30.02] & 
[29.26-29.60]        & [30.34-30.63]  & [30.62-30.89]\\
$x_{in}$                    & [0.004-0.009]  & [8$~ 10^{-4}$-0.002] & [0.003-0.007]&
[7$~ 10^{-4}$-0.001] & [0.003-0.005]  & [0.004-0.007]\\
$x_{out}$                   & [0.60-1.32]    & [0.75-1.63]   & [0.60-1.06]   & 
[0.56-1.25]          & [0.54-1.09] & [0.81-1.56]  \\
$\alpha$&  $(10^{-3})$  & $(10^{-3})$  & $(10^{-4})$ & $(10^{-4})$ & 
[2.2$~10^{-4}$-0.8$~10^{-4}$] & [4.7$~10^{-4}$-1.8$~10^{-4}$]
\\
\hline

\end{tabular}
\end{center}
\caption{\small{Best-fit parameters for the T Tauri sample. The values in square 
brackets refer to different choices of the free parameter $x_Q$ ranging in the 
interval [0.4-0.9]. The values of $\alpha$ in parentheses are {\it assumed} (see 
discussion in Sect. \ref{ttau}).}}
\label{tab:fitpar}
\end{table*}

\begin{table*}[hbt!]
\begin{center}
\begin{tabular}{ccc|cc|cc}
\hline
\hline
        & \multicolumn{2}{c}{$\alpha=10^{-3}$} & \multicolumn{2}{c}{$\alpha=
10^{-4}$} & & \\
\hline
	& BP Tau & DE Tau & BP Tau & DE Tau & DR Tau &  DG Tau\\
\hline
$\dot{M}/(10^{-6}M_{\odot}/$yr) & [3-7] & [0.81-1.7] & [0.5-0.7] & [0.12-0.26] & 
[0.5-0.5] & [1-1] \\
$M_{disk}/M_{\odot}$            & [0.28-0.44] & [0.19-0.31] & [0.28-0.3] & 
[0.13-0.21] & [0.25-0.33] & [0.41-0.5]\\
$r_s/$AU                        & [15-9]      & [20-12]     & [11-7]    & 
[16-10]     & [20-10] & [19-11]  \\
$L_{\star}/L_{\odot}$           & 0.51        & 0.44        & 0.51       & 
0.44        & 0.95  & 0.82      \\
$a/10^{-1}$                     & [0.06-0.02]   & [0.3-0.2]      & [0.4-0.1]    & 
[3.5-1.5]       & [1-1] & [0.4-0.4] \\
$\cos\theta$			& [0.03-0.02] & [0.14-0.08] & [0.12-0.07]&
[0.53-0.31] & 1 & 1\\

\hline

\end{tabular}
\end{center}
\caption{\small{Derived parameters for the disks of the T Tauri sample stars.}}
\label{tab:derpar}
\end{table*}


\begin{thebibliography}{}

\bibitem[1988]{shu}
  Adams, F.C., Lada, C.J., Shu, F.H., 1988, ApJ, 326, 865
\bibitem[1989]{adams2}
  Adams, F.C., Ruden, S.P., Shu, F.H., 1989, ApJ, 347, 959  
\bibitem[2001]{armitage}
  Armitage, P. J., Livio, M., Pringle, J. E., 2001, MNRAS, in press
\bibitem[1998]{bardou}
  Bardou, A., Heyvaerts, J., Duschl, W.J., 1998, A\&A, 337, 966
\bibitem[1990]{beckwith}
  Beckwith, S.V.W. et al., 1990, AJ, 99, 924
\bibitem[1995]{bell}
  Bell, K.R., et al., 1995, ApJ, 444, 376
\bibitem[1997]{bertin}
  Bertin, G., 1997, ApJ, 478, L71
\bibitem[1999]{lodato}
  Bertin, G., Lodato, G., 1999, A\&A, 350, 694 (BL99)
\bibitem[2001]{lodato2}
  Bertin, G., Lodato, G., 2001, A\&A, 370, 342 (BL01)
\bibitem[1996]{burrows}
  Burrows, C.J., et al., 1996, ApJ, 473, 437
\bibitem[1992]{calvet2}
  Calvet, N. et al., 1992, Rev. Mex. Aston. Astrofis., 24, 27
\bibitem[1994]{calvet}
  Calvet, N. et al., 1994, ApJ, 434, 330
\bibitem[1989]{cardelli}
  Cardelli, J.A., Clayton, G.C., Mathis, J.S., 1989, ApJ, 345, 245
\bibitem[1994]{cesaroni}
  Cesaroni, R., et al., 1994, ApJ, 435, L137
\bibitem[1997]{chiang}
  Chiang, E. I., Goldreich, P., 1997, ApJ, 490, 368
\bibitem[2000]{dullemond}
  Dullemond, C. P., 2000, A\&A, 361, L17 
\bibitem[2000]{duschl}
  Duschl, W. J., Strittmatter, P. A., Biermann, P. L., 2000, A\&A, 357, 1123
\bibitem[1998]{gullbring}
  Gullbring, E., Hartmann, L., Brice\~no, C., Calvet, N., 1998, ApJ, 492, 323
\bibitem[2000]{gullbring2}
  Gullbring, E., Calvet, N., Muzerolle, J., Hartmann, L., 2000, ApJ, 544, 927
\bibitem[1995]{hartigan}
  Hartigan, P., Edwards, S., Ghandour, L., 1995, ApJ, 452, 736 
\bibitem[1998]{hartmann} 
  Hartmann, L., 1998, ``Accretion processes in star formation'', Cambridge 
University Press, Cambridge
\bibitem[1996]{hart}
  Hartmann, L., Kenyon, S.J., 1996, ARA\&A, 34, 207
\bibitem[1971]{hohl}
  Hohl, F., 1971, ApJ, 168, 343
\bibitem[1973]{hohl2}
  Hohl, F., 1973, ApJ, 184, 353
\bibitem[1987]{kenyon2}
  Kenyon, S.J., Hartmann, L., 1987, ApJ, 323, 714
\bibitem[1991]{kenyon3}
  Kenyon, S.J., Hartmann, L., 1991, ApJ, 382, 664
\bibitem[1995]{kenyon}
  Kenyon, S.J., Hartmann, L., 1995, ApJs, 101, 117
\bibitem[1988]{hewett}
  Kenyon, S.J., Hartmann, L., Hewett, R., 1988, ApJ, 325, 231
\bibitem[1994]{kenyondr}
  Kenyon, S.J., et al., 1994, AJ, 107, 2153
\bibitem[1994]{laughlin}
  Laughlin, G., Bodenheimer, P., 1994, ApJ, 436, 335
\bibitem[1996]{laughlin2}
  Laughlin, G., R\'{o}\.{z}yczka, M., 1996, ApJ, 456, 279
\bibitem[1994]{lay}
  Lay, O. P., et al., 1994, ApJ, 434, L75
\bibitem[1987]{lin}
  Lin, D.N.C., Pringle, J.E., 1987, MNRAS, 225, 607
\bibitem[1990]{lin2}
  Lin, D.N.C., Pringle, J.E., 1990, ApJ, 358, 515
\bibitem[1974]{lynden}
  Lynden-Bell, D., Pringle, J.E., 1974, MNRAS, 168, 603
\bibitem[1998]{nelson1}
  Nelson, A.F., et al., 1998, ApJ, 502, 342
\bibitem[2000]{nelson2}
  Nelson, A.F., et al., 2000, ApJ, 529, 357
\bibitem[1998]{pickett1}
  Pickett, B.K., et al., 1998, ApJ, 504, 468
\bibitem[2000]{pickett2}
  Pickett, B.K., et al., 2000, ApJ, 529, 1034
\bibitem[1996]{popham}
  Popham, R., Kenyon, S.J., Hartmann, L., Narayan, R., 1996, ApJ, 473, 422
\bibitem[1997]{richling}
  Richling, S., Yorke, H.W., 1997, A\&A, 327, 317
\bibitem[1991]{ruden}
  Ruden, S.P., Pollack, J.B., 1991, ApJ, 375, 740
\bibitem[1976]{strom}
  Rydgren, A.E., Strom, S.E., Strom, K.M., 1976, ApJS, 30, 307
\bibitem[1973]{shakura}
  Shakura, N.J., Sunyaev, R.A., 1973, A\&A, 24, 337
\bibitem[1987]{lizano}
  Shu, F. H., Adams, F. C., Lizano, S., 1987, ARA\&A, 25, 23
\bibitem[1998]{stapelfeldt}
  Stapelfeldt, K.R., et al., 1998, ApJ, 502, L65
\bibitem[2001]{natta}
  Testi, L., Natta, A., Sheperd D. S., Wilner, D. J., 2001, preprint 
  (astro-ph/0102473)
\bibitem[1993]{valenti}
  Valenti, J. A., Basri, G., Johns, C. M., 1993, AJ, 106, 2024
\bibitem[1992]{weaver}
  Weaver, W. B., Jones, G., 1992, ApJS, 78, 239
\bibitem[2000]{wilner}
  Wilner, D.J. et al., 2000, ApJ, 534, L101
\bibitem[1999]{yorke}
  Yorke, H.W., Bodenheimer P., 1999, ApJ, 525, 330
\end{thebibliography}
\end{document}